\documentclass[aps,amsmath,amssymb,showpacs,showkeys]{revtex4-2}
\usepackage[dvips]{graphicx}
\usepackage{times}
\usepackage{braket}
\usepackage{xcolor}
\usepackage{adjustbox}
\usepackage{orcidlink}
\usepackage{hyperref}
\hypersetup{
  colorlinks=true,
  urlcolor=blue,
  linkcolor=red,
  citecolor=red
}
\begin{document}
\title{Modelling Cosmic Rays Flux with Pierre Auger and Telescope Array Data in 
$\boldsymbol{f(R)}$ and $\boldsymbol{f(Q)}$ Theories of Gravity}

\author{Swaraj Pratim Sarmah\orcidlink{0009-0003-2362-2080}}
\email[Email: ]{rs\_swarajpratimsarmah@dibru.ac.in}

\author{Umananda Dev Goswami\orcidlink{0000-0003-0012-7549}}
\email[Email: ]{umananda@dibru.ac.in}

\affiliation{Department of Physics, Dibrugarh University, Dibrugarh 786004, 
Assam, India}

\begin{abstract}
We investigate the effects of the magnetic field and the redshift on the 
propagation of galactic and extragalactic cosmic rays (CRs) in a modified 
theory of gravity (MTG) and an alternative theory of gravity (ATG) framework. 
For this purpose, we consider the $f(R)$ gravity and the $f(Q)$ gravity 
theories. We utilise these two MTG and ATG to compute the density enhancement 
factor of CRs as a function of the magnetic field and the redshift. For this 
work, we take the magnetic field strength from $1-100$ nG, while $0-2.5$ for 
the redshift. For each of these parameters, we take $100$ bins within their
considered range for the computation. The enhancement parameter for the mixed 
composition of heavy nuclei up to Fe is also taken into account for this work. 
Further, we compute the $E^3$ magnified diffusive flux of ultra high-energy
cosmic rays (UHECRs) for $150$ sources separated by a distance $d_\text{s}$ 
for the different cosmological models. For the fitting with observational data 
from the Pierre Auger Observatory (PAO) and Telescope Array (TA), we 
parameterized some set that consists of the redshift $z$, the separation 
distances $d_\text{s}$ between the $150$ sources, and the maximum cutoff 
energy $E_{\text{max}}$. For each case, a residue plot and $\chi^2$ value 
are also added to check the goodness of fit. A comparative analysis of 
the considered models has been performed in each of the 
cases along with the $\Lambda$CDM model. The $f(Q)$ model shows the 
highest CR density enhancement and the lowest reduced $\chi^2$ value when 
fitted to PAO and TA data of the UHECRs flux. The uncertainty calculations in 
flux have been performed in this work with PAO and TA data, also supporting 
the validation of the considered MTG and ATG in UHECR studies.
\end{abstract}


\maketitle                                                                      

\section{Introduction}
Cosmic rays (CRs) are ionising particles that come from the outside of our 
solar system. V.~F.~Hess discovered them in 1912 \cite{Hess}, which was a 
breakthrough for modern physics. However, even after more than a hundred and 
ten years, many questions about CRs remain unanswered 
\cite{harari, molerach, berezinskyGK}. For example, we don't know how they 
are created, accelerated and propagated, especially when they are endowed 
with high energies ($E\geq 0.1$ EeV; $1$ EeV $= 10^{18}$ eV) 
\cite{Bhattacharjee, Olinto}. We assume that the sources of lower energy 
CRs ($E \leq 0.1$ EeV) are galactic 
and are related to supernova explosions 
\cite{blasi201321, berezhko661, hewitt2015, s.mollerach}, but the sources of
higher energy CRs ($\sim 1$ EeV and above) are probably outside our galaxy 
i.e.~extragalactic \cite{Auger2017a}, and could be linked to $\gamma$-ray bursts and active 
galactic nuclei, see e.g. \cite{biermann_2012, keivani, bell, mollerach2024, marafico, mbarek} for recent studies.

The energy spectrum of CRs has an extraordinarily wide range. It extends over 
many orders of magnitude from GeV energies up to $100$ EeV and exhibits a 
power-law spectrum. There is a small spectral break known as the knee at 
about $4$ PeV ($1$ PeV $= 10^{15}$ eV) and a flattening at the so-called ankle 
at about $5$ EeV. In this spectrum, a strong cutoff near $50$ EeV, which is 
called the Greisen-Zatsepin-Kuzmin (GZK) cutoff \cite{Greisen, Zatsepin} 
appears due to the interaction of CRs with the cosmic microwave background (CMB) 
photons. At ultra-high energies (UHE), the total intensity appears as the succession of individual spectra from nuclear elements getting suppressed as the energy attains a maximum in proportion, to first order, to the charge $Z$ of the elements. The observed suppression of the all-particle spectrum \cite{harari_dark, hires2008, Auger2008, TA2013L1} thus results from both GZK effect \cite{Greisen, Zatsepin} and the maximum acceleration energy $E^{\mathrm{max}}_{Z}\simeq (5\times Z)~$ EeV of the sources \cite{PAO_jcap2017038erratum_jcap}.

The intergalactic medium (IGM) contains turbulent magnetic fields (TMFs) that 
play a crucial role in the propagation of UHECRs from extragalactic sources. 
When charged particles move through a random magnetic field, their propagation 
depends on the distance travelled relative to the scattering length, 
$l_{\text{D}} = 3D/c$, where $D$ is the diffusion coefficient and $c$ is the 
speed of light \cite{Supanitsky}. If the distance travelled by the particle is 
much shorter than the scattering length, the motion is ballistic, and if the 
travel distance is much greater, the motion becomes diffusive. Including the 
effects of extragalactic TMFs and the finite density of sources in UHECRs 
propagation studies can reveal a low-energy magnetic horizon effect 
\cite{horizon_jcap}. This effect could reconcile observations with a higher 
spectral index \cite{s.mollerach, mollerach2013, wittkowski}, closer to the 
values predicted by diffusive shock acceleration. Another hypothesis suggests 
that heavy nuclei are accelerated by extragalactic sources and then they 
interact with infrared radiation, which leads to photodisintegration and 
produces secondary nucleons, which may explain the light composition 
observed below the ankle \cite{unger, globus}. The propagation of UHECRs in 
intergalactic magnetic fields can be studied using the Boltzmann transport 
equation or various simulation methods. In Ref.~\cite{Supanitsky}, a set of 
partial differential equations is introduced to describe UHECR propagation in 
random magnetic fields, derived from the Boltzmann transport equation. The 
study highlights the diffusive nature of CR propagation. An analytical 
solution to the diffusion equation for CRs in an expanding Universe is 
provided in Ref.~\cite{berezinkyGre}, while Ref.~\cite{harari} offers a 
numerical fit for the diffusion coefficient $D(E)$ for both Kolmogorov and 
Kraichnan turbulence. The effects of CR diffusion in the local 
supercluster's magnetic field on UHECRs from nearby extragalactic sources 
are explored in Ref.~\cite{molerach}, where the authors find that a strong 
enhancement of the flux at certain energy ranges could help to explain the 
features of the CRs spectrum and composition. 
Ref.\ \cite{aloisio} provides a comprehensive analytical study of UHE 
particle propagation in extragalactic 
magnetic fields by solving the diffusion equation while considering the energy 
losses. Additionally, Ref.\ \cite{berezinsky_modif} examines the ankle, 
instep, and GZK cutoff in terms of the modification factor that arises from 
the various energy losses experienced by CR particles as they travel through 
complex galactic or intergalactic spaces \cite{berezinskyGK}. Similarly, 
Ref.~\cite{berezinski_four_feat} identifies four key features in the CRs 
proton spectrum — the ankle, instep, second ankle, and GZK cutoff by 
considering extragalactic proton interactions with the CMB and assuming a 
power-law spectrum.

General relativity (GR) is one of the most beautiful, well-tested and successful theories within the field of physics developed by Albert Einstein 
in 1915 in order to explain gravitational interactions. The theory received 
very strong backing with the detection of gravitational waves (GWs) by the 
LIGO detectors in 2015 \cite{ligo}, almost 100 years after their prediction 
from the theory by Einstein himself, and the publication of the first picture 
of the supermassive black hole which is at the heart of the M87 galaxy by the 
Event Horizon Telescope (EHT) in 2019 
\cite{m87a, m87b, m87c, m87d, m87e, m87f}. 
Some of these accomplishments, and others, have indeed buttressed the breadth 
of the GR centenary. Nevertheless, GR encounters serious difficulties 
quantitatively as well as qualitatively. As an example, gravity is yet 
integrated into a plausible frame of its quantum version. From an 
observational standpoint, GR fails to explain the present-day observations of 
the accelerating Universe \cite{reiss, perlmutter, spergel, astier} with  
dark energy \cite{Cognola:2007zu, sami, udg_prd, Odintsov:2020nwm, 
Odintsov:2019evb} and especially the rotation curves of galaxies, which point 
towards a missing mass problem \cite{Oort, Zwicky1, Zwicky2, Garrett, nashiba1}.For these and other similar problems, modified theories of gravity (MTGs) as 
well as alternative theories of gravity (ATGs) have been formulated and are 
gaining solid ground, which can explain these phenomena more satisfactorily. 

The $f(R)$ gravity theory is one of the simplest yet significant and widely 
used MTGs. This theory extends the Einstein-Hilbert (E-H) action by replacing 
the Ricci scalar $R$ with a function $f(R)$ of $R$ \cite{Sotiriou}. Many 
models related to the $f(R)$ gravity have been proposed from different 
perspectives. The most famous and viable models are the Starobinsky model 
\cite{starobinsky, staro}, Hu-Sawicki model \cite{husawicki}, Tsujikawa model 
\cite{tsujikawa}, power-law model \cite{powerlaw} and a new model reported 
recently in Ref.~\cite{gogoi_model}. The symmetric teleparallel gravity and 
its generalisation, $f(Q)$ gravity, can also be considered as a modification 
of the usual teleparallel gravity representation of ATGs, in which a
generic function $f(Q)$ of nonmetricity scalar $Q$ replaces the torsion 
scalar \cite{jimenez, harko, sanjay1, sanjay2, noemi}. $f(Q)$ gravity has been 
used to explore cosmological implications, including the behaviour of dark 
energy, and has shown potential in addressing cosmological tensions. Given the significant role of MTGs and ATGs in recent cosmological 
\cite{harko, sanjay2, gogoi_model, nashiba1, gayatri} and astrophysical 
research \cite{jbora, ronit1, ronit_scripta, nashiba2, bidyut2024}, their 
application to UHECR studies, particularly in understanding UHECR flux 
presents a promising frontier. Various research groups have studied the 
anisotropy \cite{grapes3, globus2019, Yoshiguchi_2003, mollerach2022, 
mollerech2022b, Auger2017a, Abeysekara, merksch, m.ahlers, harari2021, 
erdogdu} and propagation mechanisms \cite{mollerech2019, Sigl_1999, aloisio,
berezinsky_modif, berezinski_four_feat, berezinkyGre, prosekin} of 
UHECRs within the framework of standard cosmology, as well as through 
observations by different observatories \cite{ Auger2017, 
Auger2018, ta2019, augerprd2020, auger2022, epjweb1, biermann_2012}.

In our previous studies, we addressed the impact of $f(R)$ gravity on the 
propagation \cite{swaraj1} and anisotropy \cite{swaraj2, swaraj5} of UHECRs in a 
single-source system. We also pursued a study on the magnetic suppression of 
CR flux for many sources and various cases with mixed nuclei compositions 
considering the $f(R)$ and $f(Q)$ gravity models \cite{swaraj3} in comparison
with the result obtained from the standard $\Lambda$CDM model. In this work, 
we study, within the MTGs and ATGs paradigms, the effects of magnetic fields, 
redshift, and source distances on the propagation of both galactic and 
extragalactic CRs. Then we calculate for the first time the magnified $E^3$ 
diffusive flux for 150 sources within the frameworks of the $f(R)$ and $f(Q)$ gravity cosmological models in comparison with the results of the $\Lambda$CDM 
model. We fit a set of parameters that includes redshift and the distance 
between those $150$ sources, in such a way as to best fit our results with 
observational data from the Pierre Auger Observatory (PAO) and the Telescope 
Array (TA). We provide for each scenario a residual plot to assess the 
goodness of fit.

The structure of this work is laid out in the following form: In Section 
\ref{secII}, we dive into the quite complicated dynamics of CRs diffusion 
within a turbulent magnetic field. Section \ref{secIII} is carefully divided 
into two parts; the first part introduces the $f(R)$ power-law model, while 
the second part looks into a $f(Q)$ gravity model. In Section \ref{secIV} we 
continue the investigation of CR diffusive flux considering multicomponent 
sources and also a nuclei mixture. We leave to Section \ref{secV} the numerical 
results of our work underlining the effects due to $f(R)$ and $f(Q)$ gravity. 
We also discuss the compatibility of our results with the data provided by the 
PAO and TA along with their respective $\chi^2$ values. Finally, in Section 
\ref{secVI} we conclude with a summary and a subsequent discussion. 

\section{Diffusion of Cosmic Rays and Turbulent Magnetic Fields}\label{secII}
Due to certain limitations, modelling the extragalactic magnetic fields is 
rather challenging \cite{han}. Their exact value is still not known with 
certainty and depends on the particular region of extragalactic space 
\cite{hu_apj, urmilla}. Their field strengths at the clusters' centres change 
from a few to tens of $\mu \text{G}$ \cite{han}. In less dense locations, they 
are weaker, mostly in the range between $1$ and $10$ nG, which hints at large 
fields along the cosmic structures, such as filaments. The
magnetic fields are usually linked to the matter density. It means they are 
stronger in more dense areas like superclusters, while in voids they may be 
weaker ($\leq \sim 10^{-15}$ G). The coherence length $l_\text{c}$ is the 
largest distance over which the magnetic fields are correlated to each other.
Estimates put the magnetic field strengths of our Local Supercluster at a 
level of $1$ to $100$ nG with $l_\text{c}$ ranging from $10$ kpc up to $1$
Mpc \cite{Sigl_1999}. Within the strength of few $\mu$G, the galactic 
magnetic field (GMF) is irrelevant for the CR spectrum due to its smaller 
spatial extent, although it may have some impact on CRs' arrival direction. 
Focusing attention on the propagation 
of CRs in a turbulent and uniform extragalactic magnetic field makes our 
investigation more manageable. The defining characteristics of this field are 
its root mean square (RMS) strength $B$ and coherence length $l_\text{c}$.
$B$ is defined as $\sqrt{\langle B^2(x)\rangle}$ and varies between $1$ nG and 
$100$ nG \cite{feretti, Valle, Vazza}, while $l_\text{c}$ varies between 
$0.01$ Mpc and $1$ Mpc \cite{sigl}. Given that magnetic fields in the Local 
Superclusters are the ones with the largest impact on CRs originating from 
sources within it and are important for understanding how CRs reach Earth.

The effective Larmor radius for a charged particle with charge $Ze$ and 
energy $E$ traveling through a TMF with strength $B$ can be defined as
\begin{equation}\label{larmor} r_\text{L} = \frac{E}{ZeB} \simeq 1.1\, \frac{E/\text{EeV}}{ZB/\text{nG}}\;\text{Mpc}. \end{equation}

A crucial concept in the diffusion of charged particles in magnetic fields is 
the critical energy. This energy is defined as the energy for which the 
coherence length of a particle with charge $Ze$ is equal to its Larmor radius, 
that is $r_\text{L}(E_{\text{c}}) = l_\text{c}$. Thus, the critical energy is 
given by
\begin{equation}\label{cri_energy} E_\text{c} = ZeBl_\text{c} \simeq 0.9 Z\, \frac{B}{\text{nG}}\, \frac{l_\text{c}}{\text{Mpc}}\;\text{EeV}. \end{equation}
This critical energy distinguishes between two diffusion regimes: resonant 
diffusion at lower energies ($<E_\text{c}$) and non-resonant diffusion at 
higher energies ($>E_\text{c}$). In the resonant regime, particles experience 
significant deflections due to interactions with the magnetic field $B$ on 
scales similar to $l_\text{c}$, whereas in the non-resonant regime, 
deflections are minor and occur only over larger distances.

Numerical simulations of proton propagation have led to a formula for the 
diffusion coefficient $D$ as a function of energy \cite{harari}:
\begin{equation}\label{diff_coeff} D(E) \simeq \frac{c\,l_\text{c}}{3}\left[4 \left(\frac{E}{E_\text{c}} \right)^2 + a_\text{I} \left(\frac{E}{E_\text{c}} \right) + a_\text{L} \left(\frac{E}{E_\text{c}} \right)^{2-m} \right], \end{equation}
where $m$ is the spectral index, and $a_\text{I}$ and $a_\text{L}$ are 
coefficients. For a Kolmogorov spectrum in TMF, $m = 5/3$ with 
$a_\text{L} \approx 0.23$ and $a_\text{I} \approx 0.9$. For a Kraichnan 
spectrum, $m = 3/2$, with $a_\text{L} \approx 0.42$ and 
$a_\text{I} \approx 0.65$.

The diffusion length $l_\text{D}$, which represents the distance at which a 
particle’s overall deflection reaches about one radian, is defined by 
$l_\text{D} = 3D/c$. According to Eq.~\eqref{diff_coeff}, for 
$E/E_\text{c} \ll 0.1$, the diffusion length can be approximated as 
$l_\text{D} \simeq a_\text{L} l_\text{c} (E/E_\text{c})^{2-m}$. Conversely, 
for $E/E_\text{c} \gg 0.2$, it becomes 
$l_\text{D}  \simeq 4, l_\text{c} (E/E_\text{c})^{2}$.

In the diffusive regime, the transport equation for UHE particles 
moving through an expanding Universe, originating from a source located at 
position $x_\text{s}$, can be formulated as follows \cite{berezinkyGre}:
\begin{equation}\label{diff_eqn}
\frac{\partial \rho}{\partial t} + 3 H(t)\, \rho - b(E,t)\,\frac{\partial  \rho}{\partial E}- \rho\, \frac{\partial  \rho}{\partial E}-\frac{D(E,t)}{a^2(t)}\,\nabla^2  \rho = \frac{\mathcal{N}(E,t)}{a^3(t)}\,\delta^3({x}-{\bf{x}_\text{s}}),
\end{equation}
where $H(t)= \dot{a}(t)/a(t)$ is the Hubble parameter as a function of cosmic
time $t$, with $\dot{a}(t)$ being the time derivative of the scale factor 
$a(t)$, the coordinate ${x}$ are comoving, and particle density is $\rho$. The 
term $\mathcal{L}_\text{s}(E)$ denotes the source function describing the 
number of particles emitted with energy $E$ per unit time. At a given time 
$t$, corresponding to redshift $z$, the distance is 
$r_\text{s} = {x}-{\bf{x}_\text{s}}$. The energy losses experienced by the 
particles, due to the Universe's expansion and interactions with the CMB, are 
described by
\begin{equation}
\frac{dE}{dt} = -\, b(E,t),\;\; b(E,t) = H(t)E + b_\text{int}(E).
\end{equation}
Here, $H(t)E$ accounts for the adiabatic energy losses due to cosmic expansion, 
while $b_{\text{int}}(E)$ represents interaction losses. These interaction 
energy losses, particularly with the CMB, include processes such as pair 
production and photopion production (for details, see \cite{harari}). The 
general solution of Eq.\ \eqref{diff_eqn} was obtained in 
Ref.~\cite{berezinkyGre} as given by
\begin{equation}\label{density}
\rho(E,r_\text{s})= \int_{0}^{z_{i}} dz\, \bigg | \frac{dt}{dz} \bigg |\, \mathcal{N}(E_\text{g},z)\, \frac{\textrm{exp}\left[-r_\text{s}^2/4 \lambda^2\right]}{(4\pi \lambda^2)^{3/2}}\, \frac{dE_\text{g}}{dE},
\end{equation}
In the diffusive regime, the particle density increases based on factors such 
as energy, distance from the source, and properties of the TMF. This density 
enhancement indicates how CR density evolves due to diffusion through the 
intergalactic medium and interactions with CMB radiation \cite{swaraj1}. This 
factor can be described as the ratio of the actual particle density to the 
density that would result from rectilinear propagation, as given by 
\cite{molerach}
\begin{equation}\label{enhancement}
\xi(E,r_\text{s})=\frac{4\pi r_\text{s}^2c\, \rho(E, r_\text{s})}{\mathcal{N}(E)},
\end{equation}
The source distance $r_\text{s}$ will be modified for an ensemble of sources 
in the entire calculations, replacing it with $r_\text{i}$ which will be 
further discussed in Sec \ref{secIV}.

\section{Cosmological Models}\label{secIII}
This section introduces the cosmological models employed in calculating the 
various parameters necessary for this study. Specifically, we consider one 
model each from two gravity theories: $f(R)$ gravity and $f(Q)$ gravity. 
Additionally, we present the Hubble parameter $H(z)$ for these two models in 
the context of the aforementioned theories.

\subsection{$\mathbf{f(R)}$ Gravity Power-law Model}\label{secIIIA}
We begin by considering a simple power-law model within the framework of 
$f(R)$ gravity, expressed as \cite{udg_ijmpd, d_gogoi, powerlaw, swaraj_jop}
\begin{equation}
f(R) = \beta R^n,
\end{equation}
where $\beta$ and $n$ are model parameters. The parameter $\beta$ depends on 
$n$ and other cosmological parameters such as the Hubble constant $H_0$, the 
current matter density parameter $\Omega_{\text{m}0}$, and the radiation 
density parameter $\Omega_{\text{r}0}$ through the present-day Ricci scalar 
$R_0$ as shown in Ref.~\cite{d_gogoi}:
\begin{equation}\label{lambda}
\beta = -\,\frac{3H_0^2\, \Omega_{\text{m}0}}{(n-2)R_0^n}.
\end{equation} 
The current Ricci scalar $R_0$ is given by \cite{d_gogoi}
\begin{equation}\label{R0}
R_0 = -\, \frac{3 (3-n)^2 H_0^2\, \Omega_{\text{m}0}}{2n\left[(n-3)\Omega_{\text{m}0} + 2 (n-2) \Omega_{\text{r}0}\right]}.
\end{equation}

Using the Palatini formalism, the Friedmann equation in $f(R)$ gravity theory,
in terms of redshift $z$ can be expressed as \cite{Santos}
\begin{equation}\label{fredmann}
\frac{H^2}{H_0^2}=\frac{3\,\Omega_{\text{m}0}(1+z)^3 + 6\,\Omega_{\text{r}0}(1+z)^4 + \frac{f(R)}{H_0^2}}{6 f'(R)\zeta^2 },
\end{equation}
where
\begin{equation}\label{zeta}
\zeta = 1+ \frac{9f''(R)}{2f'(R)}\frac{H_0^2\, \Omega_{\text{m}0}(1+z)^3}{Rf'(R)-f'(R)}.
\end{equation}
From Eqs.~\eqref{fredmann} and \eqref{zeta}, the Hubble parameter as a 
function of redshift $z$ can be derived as
\begin{equation}\label{powerlawhubble}
H(z) = \left[-\,\frac{2nR_0}{3 (3-n)^2\, \Omega_{\text{m}0}} \left\{(n-3)\Omega_{\text{m}0}(1+z)^{\frac{3}{n}} + 2 (n-2)\,\Omega_{\text{r}0} (1+z)^{\frac{n+3}{n}} \right\}\right]^\frac{1}{2}\!\!\!.
\end{equation}
In this study, we adopt the value $n=1.4$, which has been identified as the 
best-fit parameter with the Hubble dataset as discussed in Ref.~\cite{d_gogoi}. The 
other cosmological parameters that we used are 
$H_0 \approx 67.4$ km s$^{-1}$ Mpc$^{-1}$ \cite{planck2018}, 
$\Omega_{\text{m}0} \approx 0.315$ \cite{planck2018}, and 
$\Omega_{\text{r}0} \approx 5.373 \times 10^{-5}$ \cite{nakamura}. The 
relationship between the cosmological time evolution and redshift can thus be 
expressed as \cite{swaraj1}
\begin{equation}\label{dtdz1}
\bigg | \frac{dt}{dz} \bigg |_{f(R)} =\frac{1}{(1+z)\, H} =(1+z)^{-1}  \left[-\,\frac{2nR_0}{3 (3-n)^2 \Omega_{\text{m}0}} \left\{(n-3)\Omega_{\text{m}0}(1+z)^{\frac{3}{n}} + 2 (n-2)\Omega_{\text{r}0} (1+z)^{\frac{n+3}{n}} \right\} \right]^{-\,\frac{1}{2}}.
\end{equation}
This equation will be employed in Section \ref{secIV} to calculate the CRs 
flux for the $f(R)$ gravity model.

\subsection{The $\mathbf{f(Q)}$ Gravity Model}\label{secIIIB}
The $f(Q)$ gravity model considered in this work is defined as \cite{solanki}
\begin{equation}
f(Q)= \sigma Q,
\end{equation}
where $\sigma$ is the model parameter. For this $f(Q)$ gravity model, the 
Hubble parameter in terms of redshift $z$ is given by \cite{solanki}
\begin{equation}\label{hzfq}
H(z) = H_0 \left[(1+z)^{\frac{3\sigma + C_1+C_2}{2 \sigma + C_2}} \left( 1+ \frac{C_0}{3\sigma + C_1+C_2} \right)- \frac{C_0}{3\sigma + C_1+C_2} \right],
\end{equation}
where $C_0$, $C_1$, and $C_2$ are constants related to bulk viscous effects. 
The best-fit values for these parameters are $\sigma= -1.03^{+0.52}_{-0.55}$, 
$C_0=1.54^{+0.83}_{-0.79}$, $C_1=0.08^{+0.49}_{-0.49}$, and $C_2=0.66^{+0.82}_{-0.83}$ as reported in Ref.~\cite{solanki}. Similar to the $f(R)$ model, the 
relation between the cosmological time evolution and redshift for this $f(Q)$ 
model is given by
\begin{equation}\label{dtdz2}
\bigg | \frac{dt}{dz} \bigg |_{f(Q)} = (H_0(1+z))^{-1} \left[(1+z)^{\frac{3\sigma + C_1+C_2}{2 \sigma + C_2}} \left( 1+ \frac{C_0}{3\sigma + C_1+C_2} \right)- \frac{C_0}{3\sigma + C_1+C_2} \right]^{-1}.
\end{equation}
This relation will be used in Section \ref{secIV} to compute the CRs flux for 
the $f(Q)$ gravity model.

\begin{figure}[htb!]
\centerline{
\includegraphics[scale=0.45]{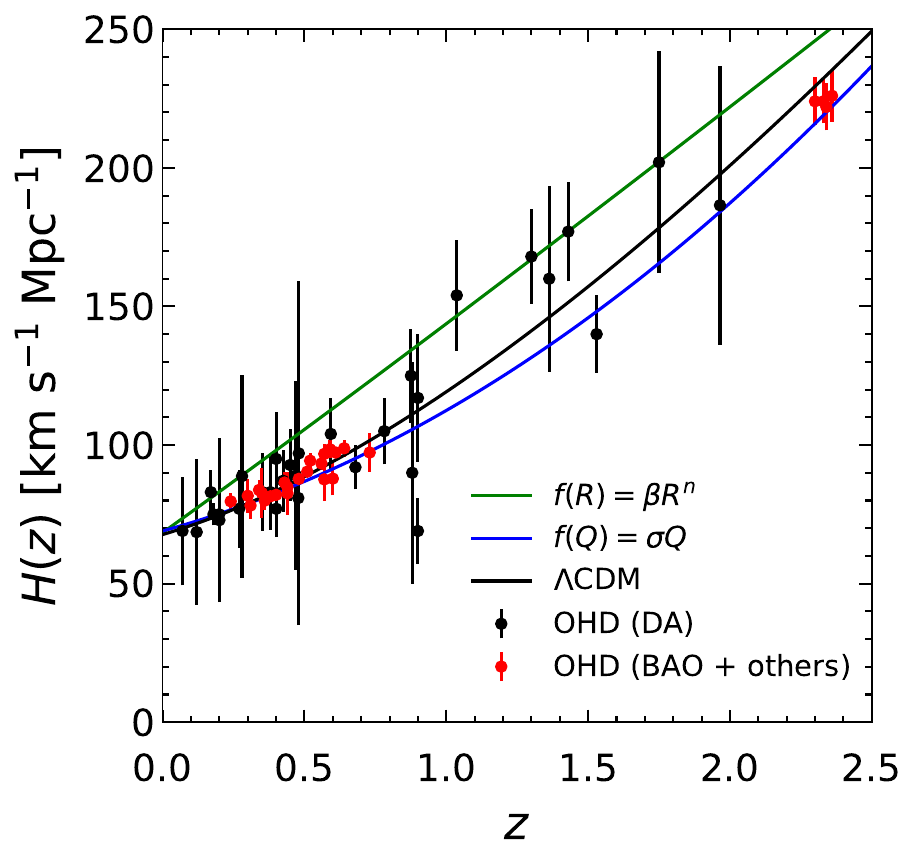}\hspace{0.5cm}
\includegraphics[scale=0.45]{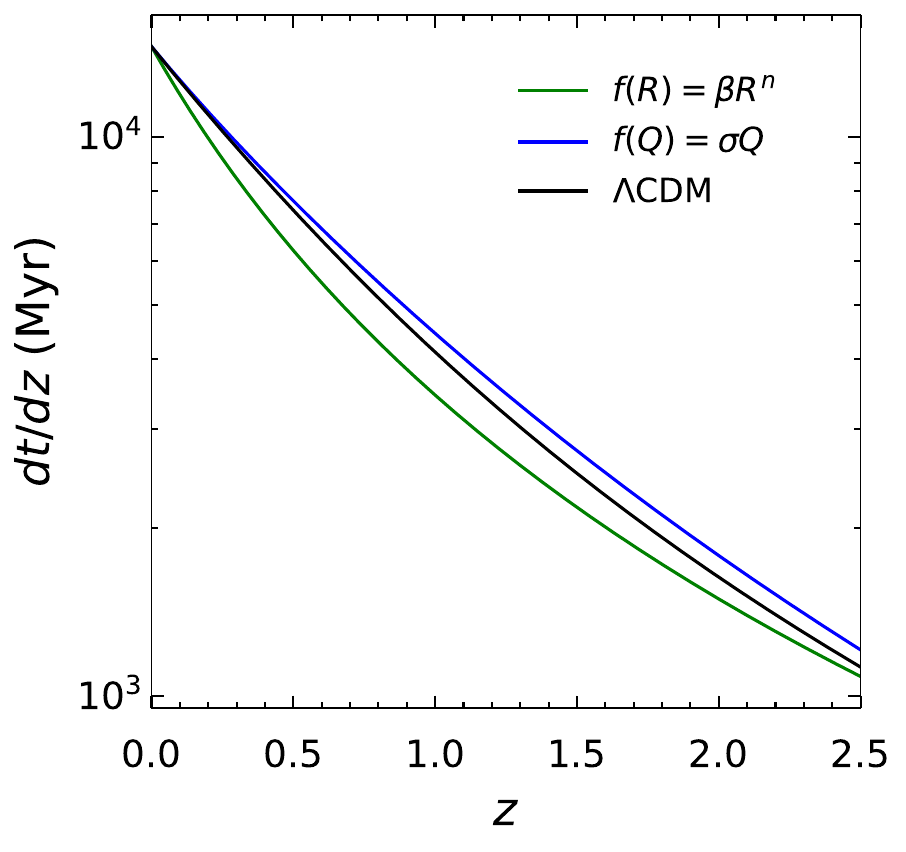}
}
\vspace{-0.2cm}
\caption{Left: Variations of Hubble parameter $H(z)$ with redshift $z$ as
predicted by the $f(R)$ gravity power-law model, $\Lambda$CDM model, and 
the $f(Q)$ gravity model in comparison with the observational Hubble data 
(OHD) obtained from differential age (DA) and Baryon Acoustic oscillations 
(BAO) methods \cite{solanki, swaraj1, d_gogoi}. Right: Evolution of 
cosmological time evolution with the redshift $z$ as predicted by the 
considered models.}
\label{fig0}
\end{figure}
In the left panel of Fig.~\ref{fig0}, we compare the Hubble parameters for the 
$f(R)$ power-law model, the $f(Q)$ gravity model, and the standard $\Lambda$CDM 
model using Eqs.~\eqref{powerlawhubble} and \eqref{hzfq}, respectively, along 
with observational Hubble data (OHD) obtained from differential age (DA) and 
Baryon Acoustic Oscillations (BAO) methods \cite{solanki, swaraj1, d_gogoi}. 
The plot demonstrates that both modified and alternative gravity models fit 
well with the observational data. It is worth noting that the models predict 
slightly different values of the Hubble constant $H_0$, with $67.77$ km 
s$^{-1}$ Mpc$^{-1}$ for the $\Lambda$CDM, $68.4$ km s$^{-1}$ Mpc$^{-1}$ for the 
$f(R)$ power-law model, and $69.0$ km s$^{-1}$ Mpc$^{-1}$ for the $f(Q)$ 
model \cite{solanki}. The values for the $\Lambda$CDM and the $f(R)$ model are 
close to the value of $H_0$ observed by the Planck experiment \cite{planck2018}.
For completeness, the right panel of Fig.~\ref{fig0} illustrates the 
cosmological time evolution with respect to redshift $z$ for all three models. 
As evident from the plot, at the 
present-day redshift ($z=0$), the predictions of the models are nearly 
indistinguishable. However, as $z$ increases, the differences become more 
significant, particularly for the $f(R)$ model. At $z=2.5$, the $\Lambda$CDM 
and $f(R)$ models converge, while the $f(Q)$ model exhibits a flat pattern.

\section{Cosmic Rays' Flux For an ensemble of sources}\label{secIV}
Several researchers have studied CRs' diffusion in the TMFs \cite{berezinkyGre, 
blasi, globus, sigl, stanev, kotera, Yoshiguchi, lemonie1, hooper, hooper2, 
sigl2007, aloisio.ptep}. Berezinsky and Gazizov \cite{9, berezinkyGre} have 
further developed the Syrovatskii solution \cite{syrovatsy_1959} in order to 
study the diffusion of protons in an expanding Universe. The flux expected from 
a CR source located at a distance $r_\text{s}$, which is much greater than 
the diffusion length $l_\text{D}$, can be calculated by solving the diffusion 
equation in the context of an expanding Universe \cite{berezinkyGre}. The 
resulting expression is given as follows \cite{Manuel}:
\begin{equation}\label{fluxeq}
J(E) = \frac{c}{4\pi} \int_{0}^{z_{\text{max}}} \!\! dz \, \left| \frac{dt}{dz} \right| \, \mathcal{N}\left[E_\text{g}(E, z), z\right] \frac{\exp\left[-r_\text{s}^2 / (4 \lambda^2)\right]}{(4 \pi \lambda^2)^{3/2}} \frac{dE_\text{g}}{dE},
\end{equation}
where $z_{\text{max}}$ denotes the maximum redshift at which the source begins 
emitting CRs, $E_\text{g}(E, z) $ is the generation energy at redshift $z$ 
corresponding to an energy  $E$ at $z = 0$, and $ \mathcal{N} $ represents 
the source emissivity obtained by summing the contributions of the 
charge-specific emissivity $\mathcal{N}_\text{Z} $ for different charges. The 
charge-specific emissivity is described by a power-law with a rigidity cutoff 
$ZE_\text{max} $, given by $ \mathcal{N}_\text{Z}(E, z) = \xi_\text{Z} f(z) E^{-\gamma} / \cosh(E / ZE_\text{max}) $ \cite{mollerach2013}, where $ \xi_\text{Z}$ 
indicates the relative contribution of charge $Z$ nuclei to the CR flux, and 
$ f(z) $ represents the evolution of the source emissivity with redshift $ z $. A different source evolution case
is discussed in Ref. \cite{swaraj6}.
The Syrovatskii variable $ \lambda^2 $ is defined as
\begin{equation}\label{syro}
\lambda^2(E, z) = \int_{0}^{z} dz \, \left| \frac{dt}{dz} \right| (1 + z)^2 D(E_g, z).
\end{equation}
Although Eq.~\eqref{fluxeq} was initially derived for protons, it is also 
applicable to nuclei when expressed in the context of particle rigidities. 
Photo-disintegration processes in nuclei generally conserve the Lorentz factor 
and rigidity of the main fragment, thus minimally affecting the particle's 
diffusion properties. However, complications arise from photo-disintegration 
losses, as the source term $ \mathcal{N} $ pertains to the primary nucleus 
leading to the observed one, which is difficult to determine due to the 
stochastic nature of the process. This scenario was previously addressed by 
S.~Mollerach et al.~\cite{mollerach2013}, and we will extend this discussion 
within modified and alternative gravity frameworks. Since our focus is on 
multiple sources rather than a single source, we use the propagation 
theorem \cite{aloisio} to sum all sources, which can be expressed as
\begin{equation}\label{lim1}
\int_{0}^{\infty} dr \, 4\pi r^2 \frac{\exp\left[-r^2 / (4 \lambda^2)\right]}{(4 \pi \lambda^2)^{3/2}} = 1.
\end{equation}
                                                       
To investigate how the finite distance to sources influences suppression, we 
calculate the sum using a specific set of distance distributions. These 
distributions assume a uniform source density, with the source distances from 
the observer given by \cite{mollerach2013, Manuel}
\begin{equation}\label{ri}
r_\text{i} = \left(\frac{3}{4\pi}\right)^{\!1/3}\!\!\! d_\text{s}\, \frac{\Gamma(i + 1/3)}{(i - 1)!},
\end{equation}
where $d_\text{s}$ represents the distance between the sources and $i$ 
indicates the $i$-th source from the average distance. Therefore, for a 
discrete source distribution, summing over the sources results in a specific 
factor \cite{mollerach2013, Manuel}
\begin{equation} \label{F_supp}
F \equiv \frac{1}{n_\text{s}} \sum_i \frac{\exp\left[-r_\text{i}^2/4 \lambda^2\right]}{(4\pi \lambda^2)^{3/2}}
\end{equation}
instead of getting Eq.~\eqref{lim1}, where $n_\text{s}$ is the source density.

In Eq.~\eqref{fluxeq}, after summing all the sources, we can write the 
modified flux for an ensemble of sources for the $f(R)$ gravity power-law 
model as
\begin{align}\label{flux_f(R)}
J_\text{mod}(E) \Big |_{f(R)} \simeq  \frac{R_\text{H}\, n_\text{s}}{4\pi}\! \int_{0}^{z_\text{max}}\!\!\!\!\! dz\, (1+z)^{-1}  &\left[-\,\frac{2nR_0}{3 (3-n)^2 \Omega_{\text{m}0}} \Bigl\{(n-3)\Omega_{\text{m}0}(1+z)^{\frac{3}{n}} + 2 (n-2)\Omega_{\text{r}0} (1+z)^{\frac{n+3}{n}} \Bigl\} \right]^{-\,\frac{1}{2}} \nonumber \\[5pt]
&\times \mathcal{N}\left[E_\text{g}(E, z), z\right]\, \frac{dE_\text{g}}{dE}\, F,
\end{align}
where $R_\text{H} = c/H_0$ is the Hubble radius. Similarly, for the $f(Q)$ 
gravity model, the modified flux can be written as
\begin{align}\label{flux_f(Q)}
J_\text{mod}(E) \Big |_{f(Q)} \simeq  \frac{R_\text{H}\, n_\text{s}}{4\pi} \int_{0}^{z_\text{max}}\!\!\!\!\!dz\,  (1+z)^{-1}  &\left[(1+z)^{\frac{3\sigma + C_1+C_2}{2 \sigma + C_2}} \left( 1+ \frac{C_0}{3\sigma + C_1+C_2} \right)- \frac{C_0}{3\sigma + C_1+C_2} \right]^{-1} \nonumber \\[5pt]
& \times \mathcal{N}\left[E_\text{g}(E, z), z\right]\, \frac{dE_\text{g}}{dE}\, F.
\end{align}
Moreover, we can rewrite Eq.~\eqref{syro} in terms of the Hubble radius 
$R_\text{H}$ and from Eq.~\eqref{diff_coeff} as
\begin{equation}\label{ad}
\lambda^2(E,z)= H_0\frac{R_\text{H} l_\text{c}}{3}\int_{0}^{z}\!\!dz\, \bigg | \frac{dt}{dz} \bigg |\,(1+z)^2 \left[4 \left(\frac{(1+z)\,E}{E_\text{c}} \right)^2 + a_\text{I} \left(\frac{(1+z)\,E}{E_\text{c}} \right) + a_\text{L} \left(\frac{(1+z)\,E}{E_\text{c}} \right)^{\alpha}   \right].
\end{equation}

Using these above relations in the framework of the considered models of MTG 
and ATG, we will discuss the numerical results for the CR density enhancement 
factor and the flux in the next section.

\section{Numerical results}\label{secV}
As mentioned earlier this section is dedicated to numerical calculations, data 
fitting and chi-square testing. We extensively use the \texttt{python scipy} 
and \texttt{numpy} library for numerical computations and \texttt{matplotlib} 
for the plotting. Unless otherwise specified, in all subsequent plots assume 
the primary particle to be a proton with a spectral index of $\gamma = 2$.

\begin{figure}[!h]
\centerline{
\includegraphics[scale=0.4]{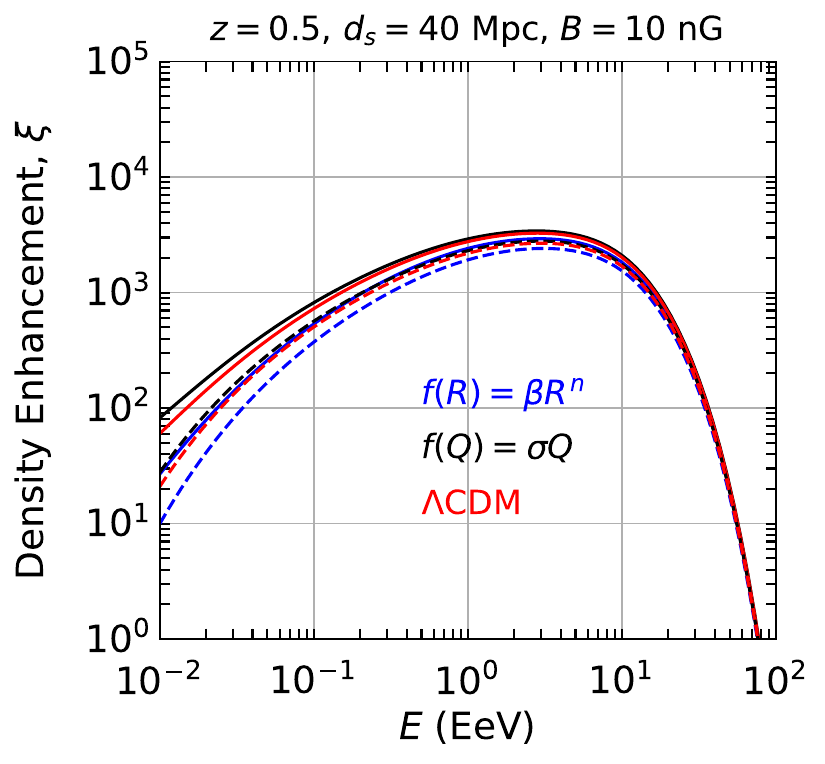}
\includegraphics[scale=0.4]{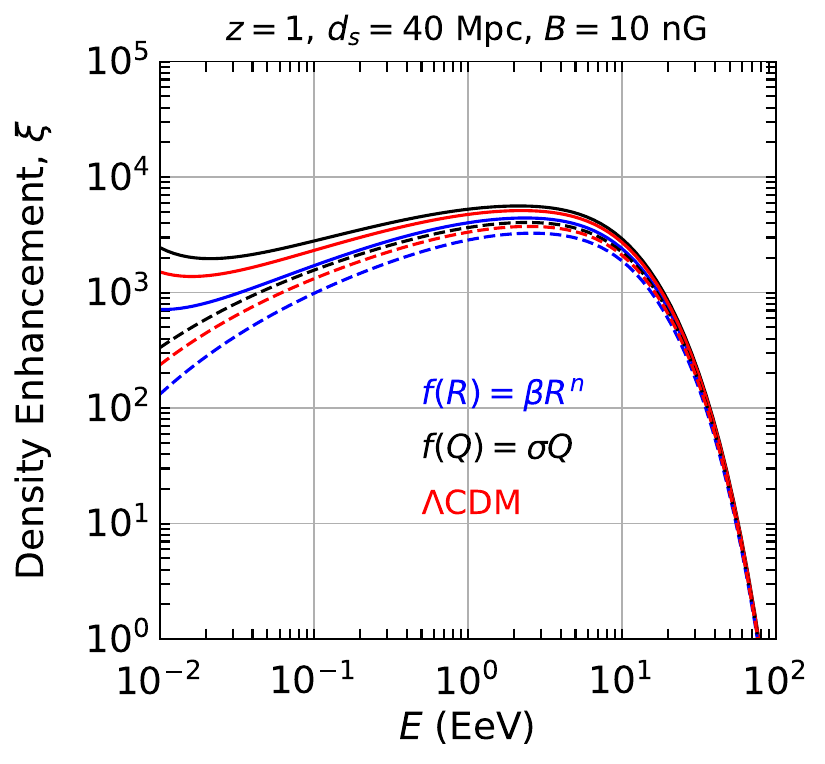}
\includegraphics[scale=0.4]{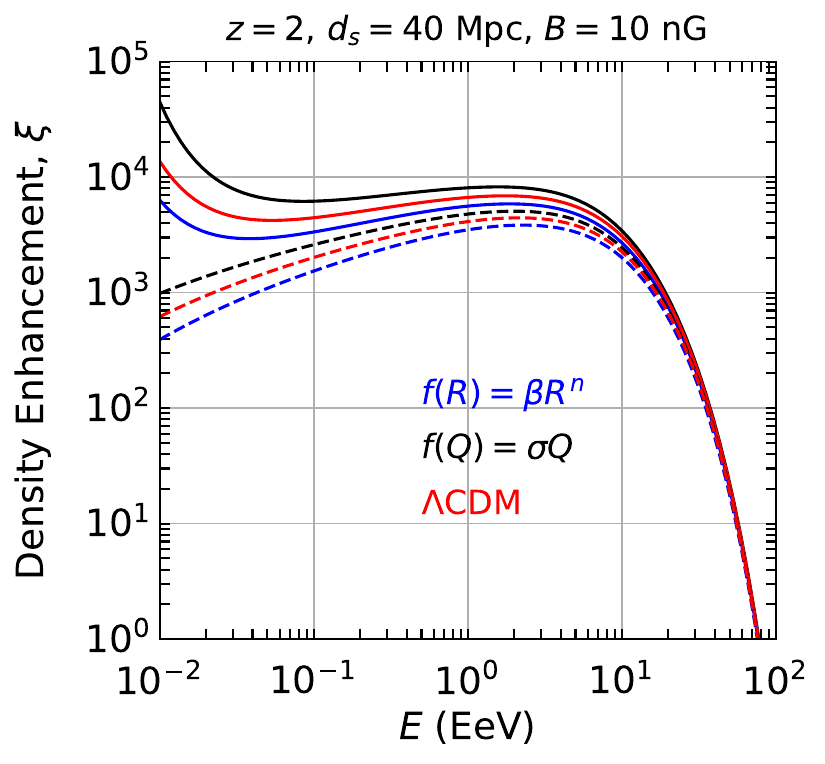}}
\vspace{-0.2cm}
\caption{The density enhancement factor $\xi$ as a function of energy $E$ as
predicted by the $f(R)$ gravity, $f(Q)$ gravity and $\Lambda$CDM models for 
$z=0.5-2$, $l_\text{c}=1$ Mpc, and $B=10$ nG. Here, the solid lines 
represent the case with energy loss processes, while the dashed lines 
correspond to the case without energy losses.}
\label{enhancement_comp1}
\end{figure}

\begin{figure}[!h]
\centerline{
\includegraphics[scale=0.375]{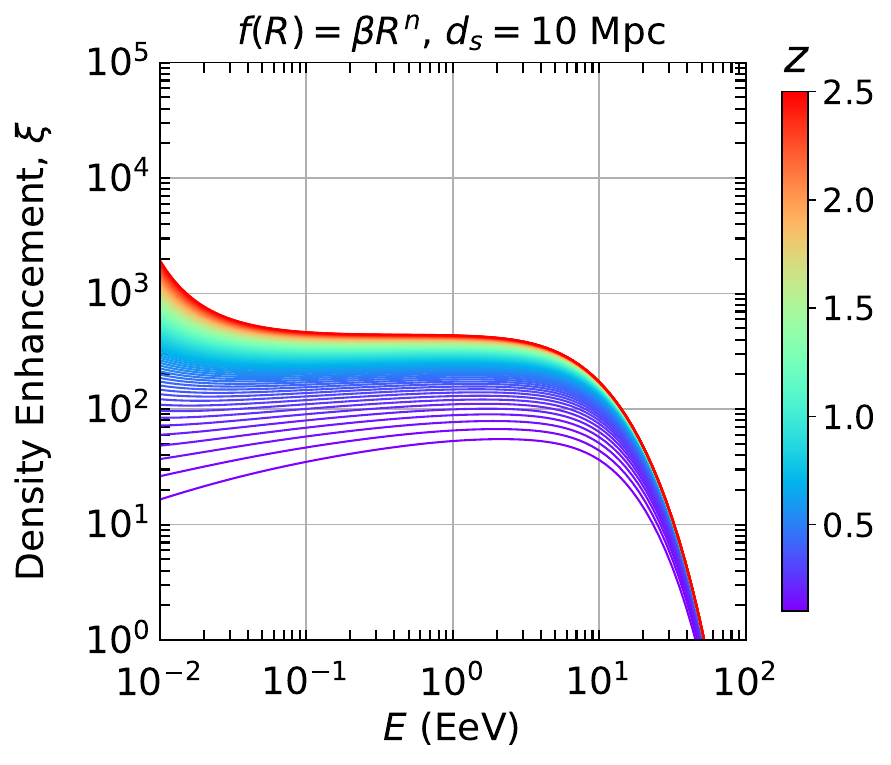}
\includegraphics[scale=0.375]{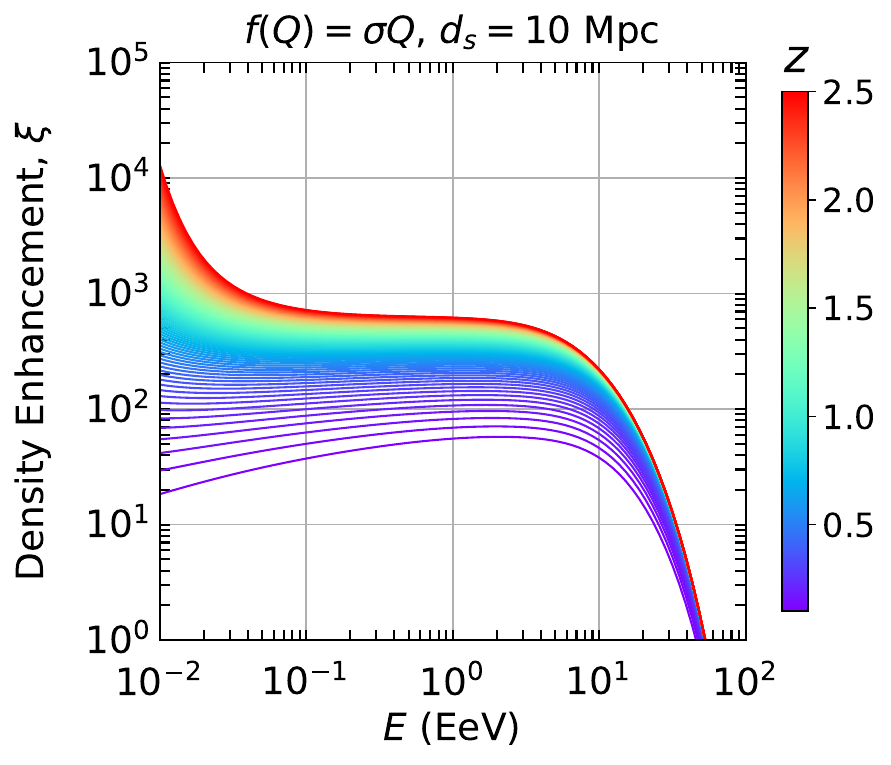}
\includegraphics[scale=0.375]{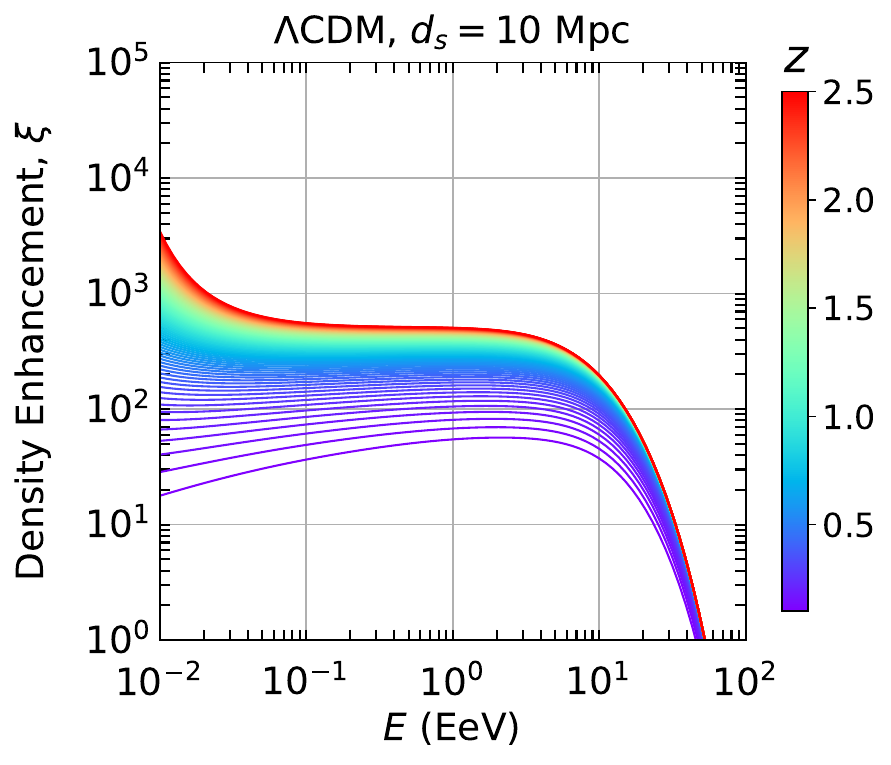}
}\vspace{0.2cm}
\centerline{
\includegraphics[scale=0.375]{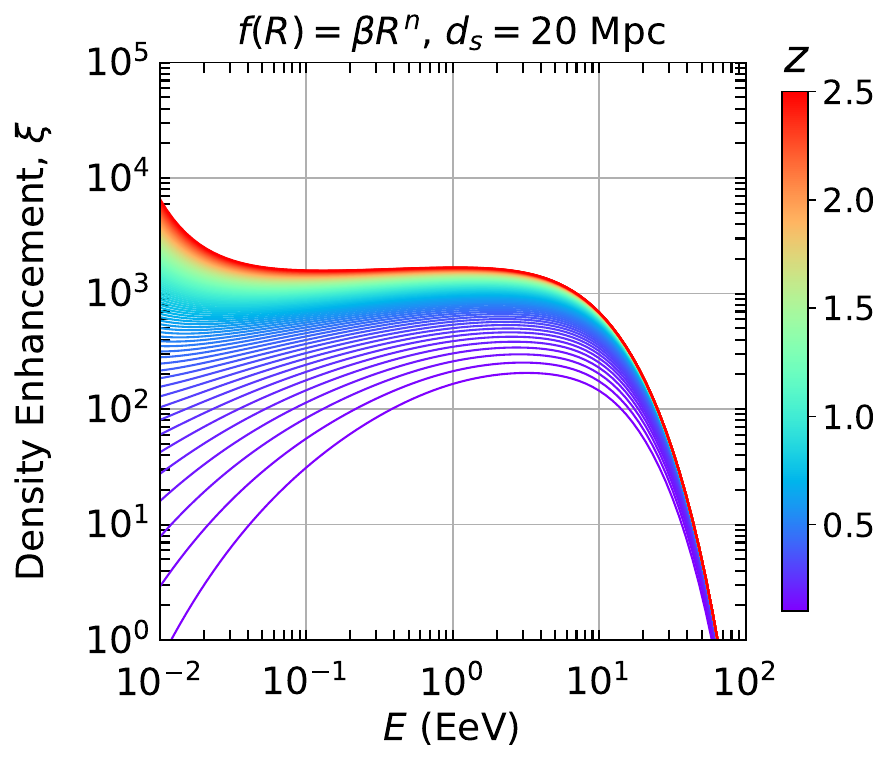}
\includegraphics[scale=0.375]{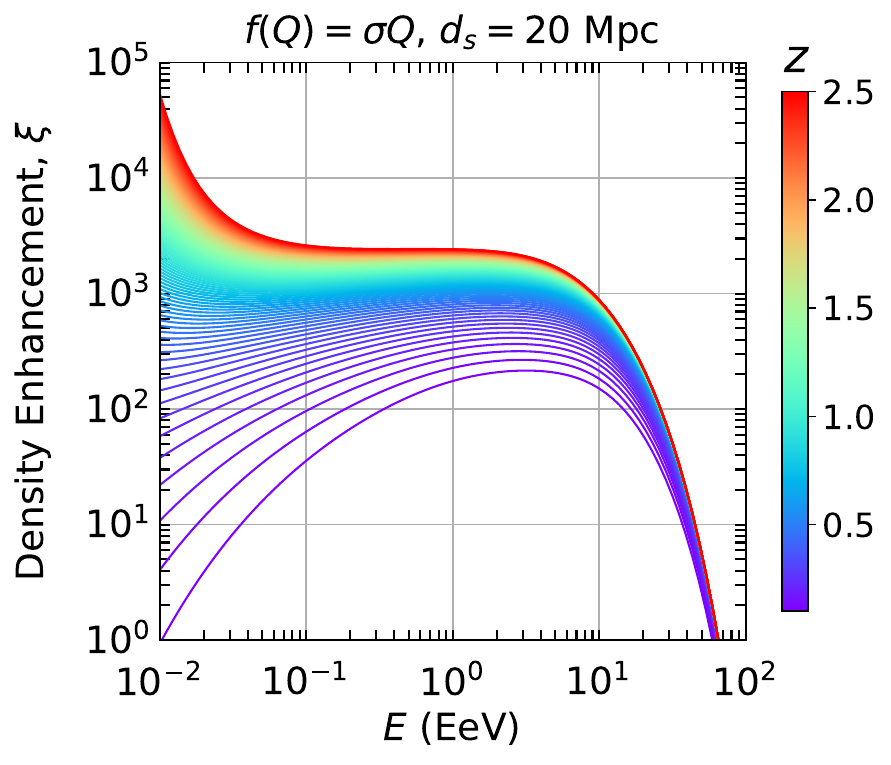}
\includegraphics[scale=0.375]{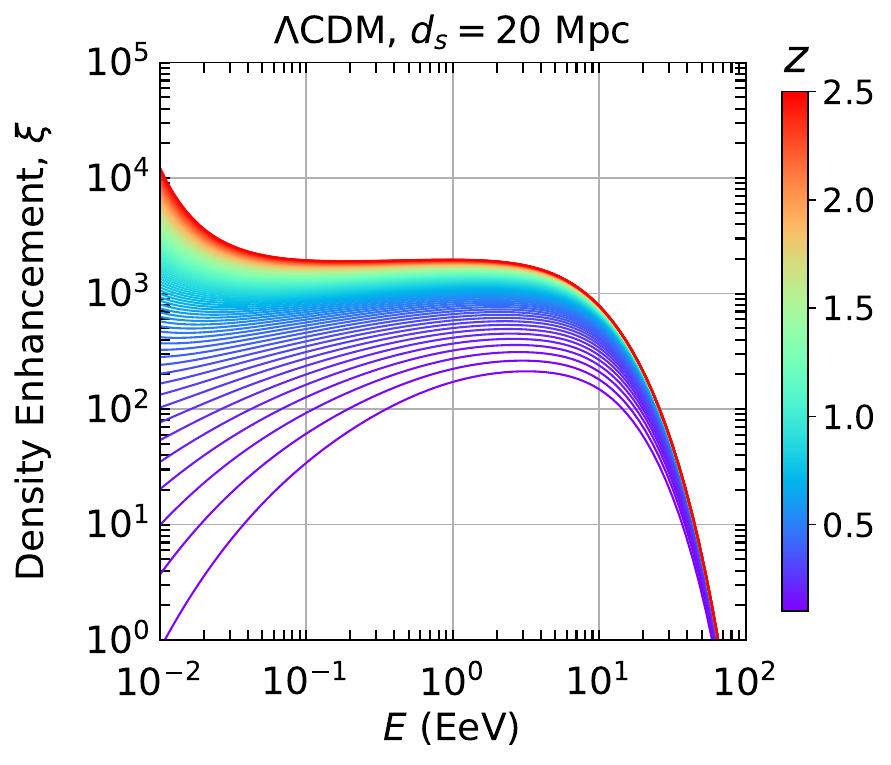}
}\vspace{0.2cm}
\centerline{
\includegraphics[scale=0.375]{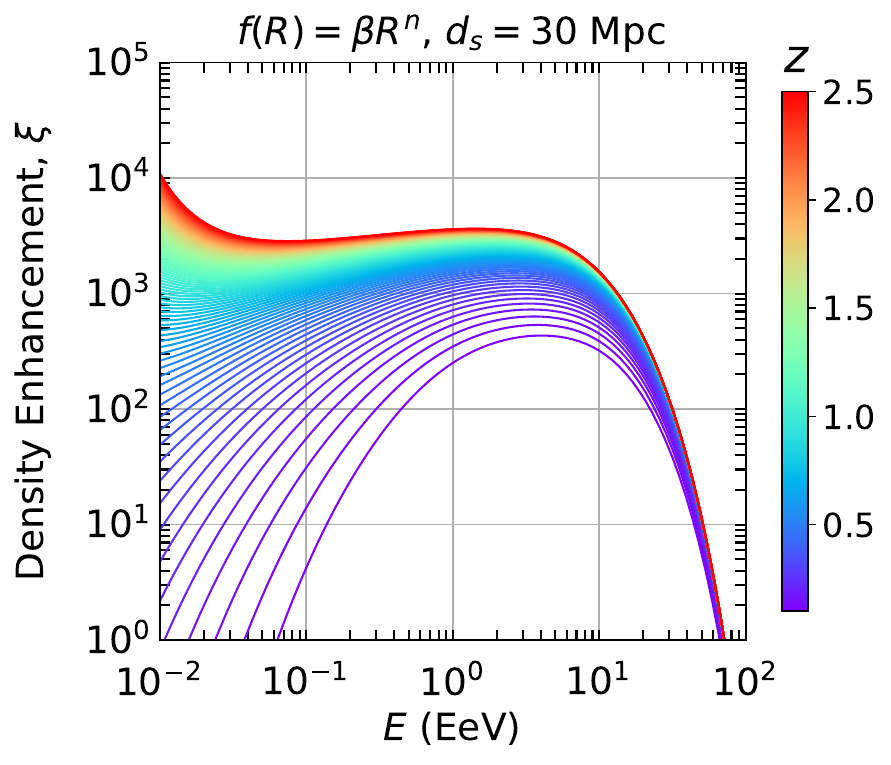}
\includegraphics[scale=0.375]{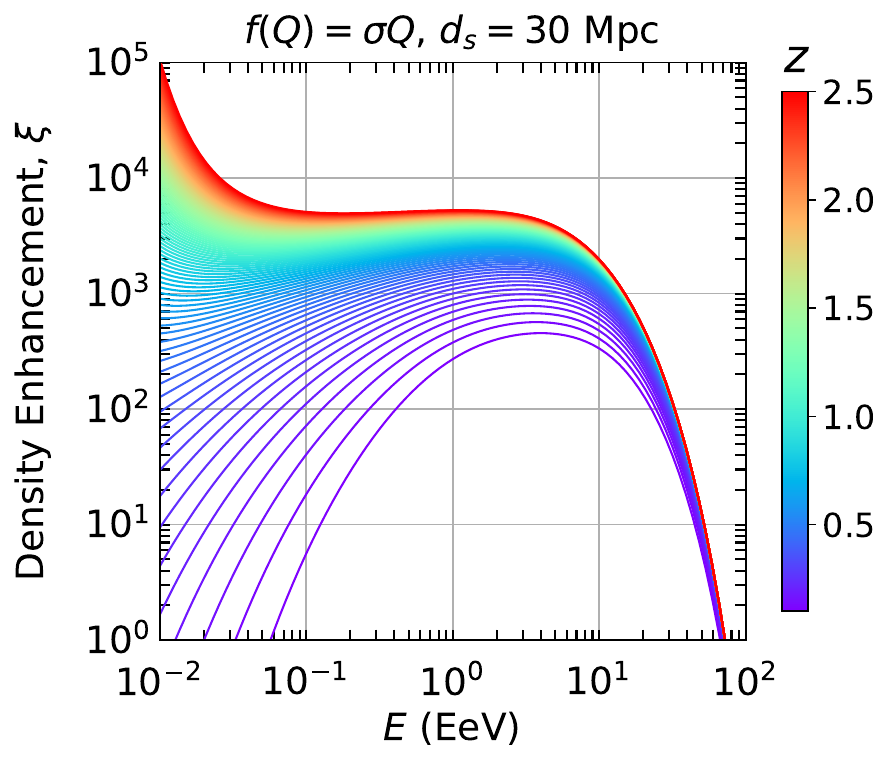}
\includegraphics[scale=0.375]{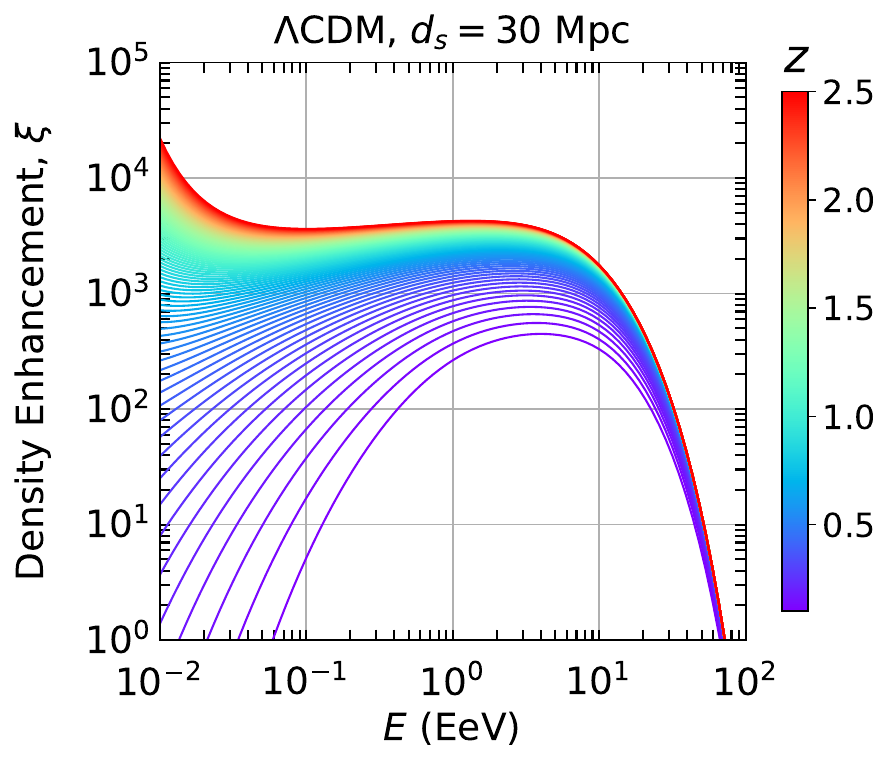}
}\vspace{0.2cm}
\centerline{
\includegraphics[scale=0.375]{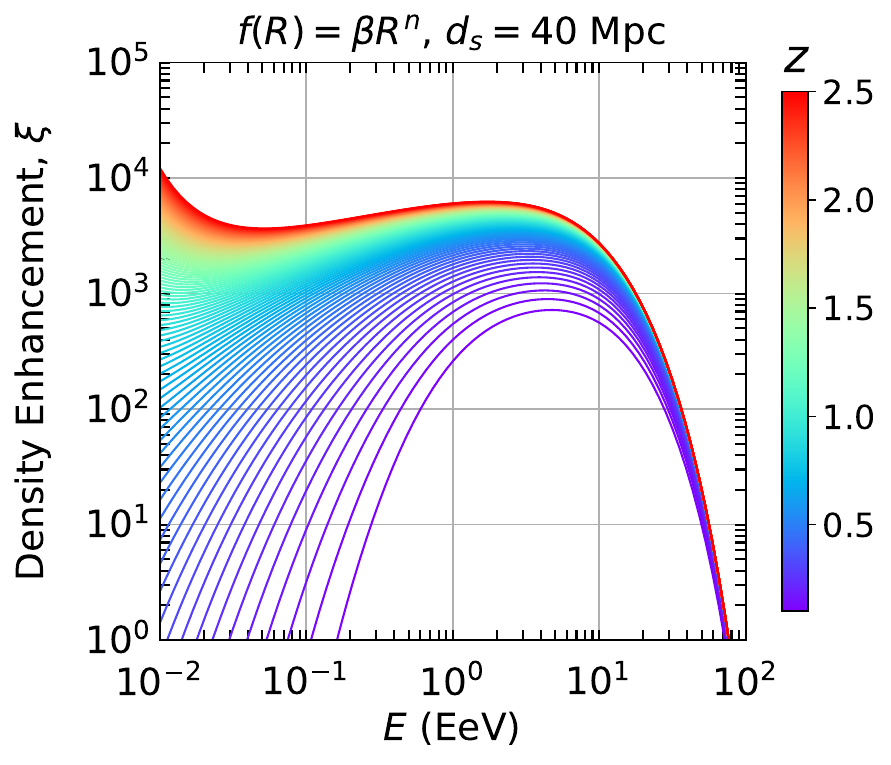}
\includegraphics[scale=0.375]{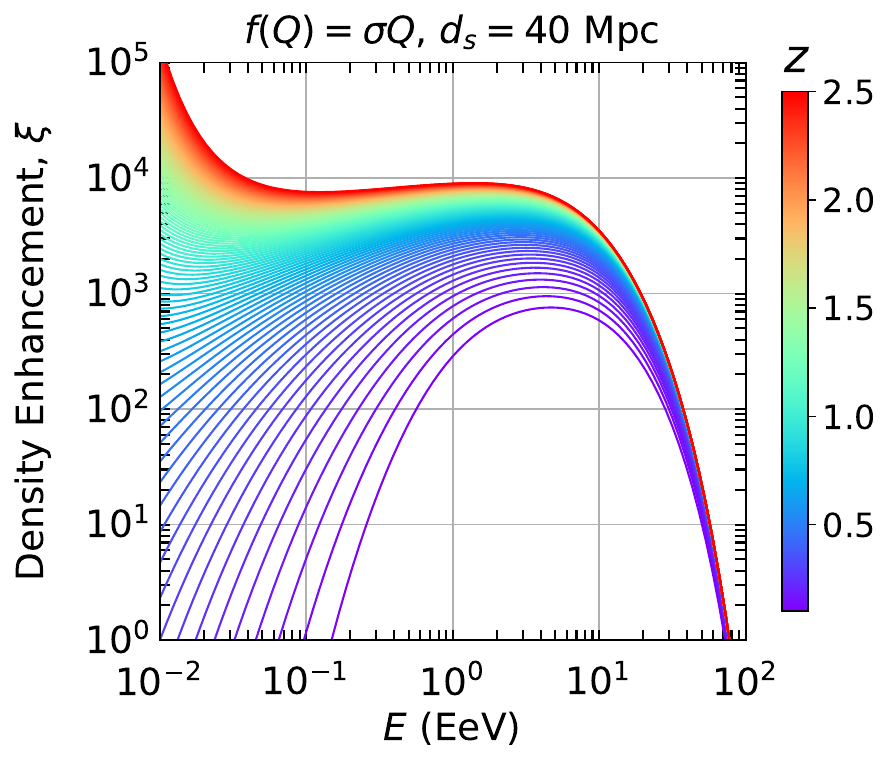}
\includegraphics[scale=0.375]{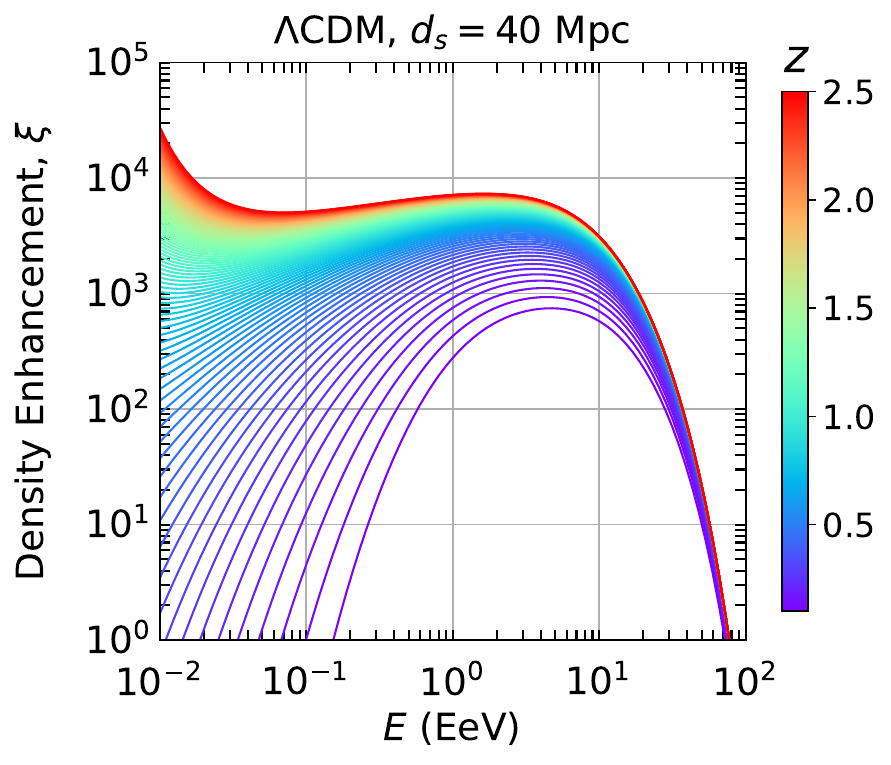}
}
\vspace{-0.2cm}
\caption{The density enhancement factor $\xi$ as a function of energy $E$ for 
$z=0-2.5$ by considering $B=10$ nG, $l_\text{c}=1$ Mpc. The left panels 
represent the results from the $f(R)$ gravity model, the middle panels 
represent the results from the $f(Q)$ gravity model, and the right panels 
represent the results from the standard $\Lambda$CDM model. The separation 
distances between the sources are $10$ Mpc, $20$ Mpc, $30$ Mpc and $40$ Mpc 
from top to bottom.}
\label{fig_enhancement_z}
\end{figure}

We compute the enhancement factor of both galactic and extragalactic CRs 
within the framework of MTG and ATG. For computational purposes, as 
already mentioned, we use two models, each of MTGs and ATGs along with the 
standard $\Lambda$CDM model, viz.~the $f(R)$ power-law model and the $f(Q)$ 
model, which are discussed in the previous section. A comparative analysis has 
been performed for the CR density enhancement factor by considering the 
variation of redshift from $z=0.5-2$, source separation distance 
$d_\text{s}=40$ Mpc and 
magnetic field strength $B=10$ nG in Fig.~\ref{enhancement_comp1}. 
For the clarity of the behaviour of the enhancement of CRs at different
energy regions, we consider here two scenarios: the solid lines represent 
results that include the energy loss interactions, while the dashed lines 
correspond to those with the assumption of without energy losses.
We can see that as the redshift has taken its higher value, the enhancement 
factor also tends to its higher amplitude, especially at the lower side of the 
primary energy of CR particles (solid lines). As it is clear from the
figure the increase in the enhancement (solid limes) in lower energies (second 
and third panel) is due to redshift related losses, which become substantial 
at higher $z$ values. 
The important point is that the cosmological model effect is 
more pronounced at higher redshift rather than lower ones and in the
lower energy side. For more clarification of these behaviours, we draw 
Fig.~\ref{fig_enhancement_z} for a range of redshifts from $0$ to $2.5$.

In Fig.~\ref{fig_enhancement_z}, we plot the density enhancement factor of CRs 
within an energy range from $0.01$ EeV up to $100$ EeV. For this scenario, we 
take the cosmological redshift from $0$ to $2.5$. We take $100$ bins in 
between the redshift range of $0$ to $2.5$ and plot the enhancement factor 
with respect to the energy $E$ for each value of the redshift bin. We consider 
the distance between the sources $d_\text{s}$ as $10$ Mpc, $20$ Mpc, $30$ Mpc 
and $40$ Mpc (top to bottom). From this Fig.~\ref{fig_enhancement_z} we observe 
that at the lower redshift values ($<1$), the density enhancement increases 
with increasing energy up to $\sim 5$ EeV then decreases after that. But at 
the higher values of redshift ($>1$), the density enhancement decreases for 
increasing energy up to $0.1$ EeV i.e. the galactic CRs region. The 
density enhancement decreases slowly with the energy in the top panel, i.e. 
$d_\text{s}=10$ Mpc. In the second panel from the top ($d_\text{s}=20$ Mpc), it 
follows a flat pattern within that energy range. At $d_\text{s}=30$ Mpc (third 
panel from top), it increases slowly, then finally at $d_\text{s}=40$ Mpc 
(bottom panel), a well-increasing pattern is visible. Thus, we can say that 
as the distances between the sources increase, the enhancement factor also 
increases. The enhancement factor is observed up to $\sim 70$ EeV, after that 
no enhancement has been observed in our study. Again, we implement the results 
with the three cosmological models. The $f(R)$ model (left panel) predicts 
the lowest enhancement, the highest enhancement by the $f(Q)$ model (middle 
panel), and the standard $\Lambda$CDM model (right panel) predicts a moderate 
enhancement in the entire energy range we consider here. The effects of the 
cosmological models are more pronounced in the higher $d_\text{s}$ values and 
higher redshifts. 

\begin{figure}[htb!]
\centerline{
\includegraphics[scale=0.4]{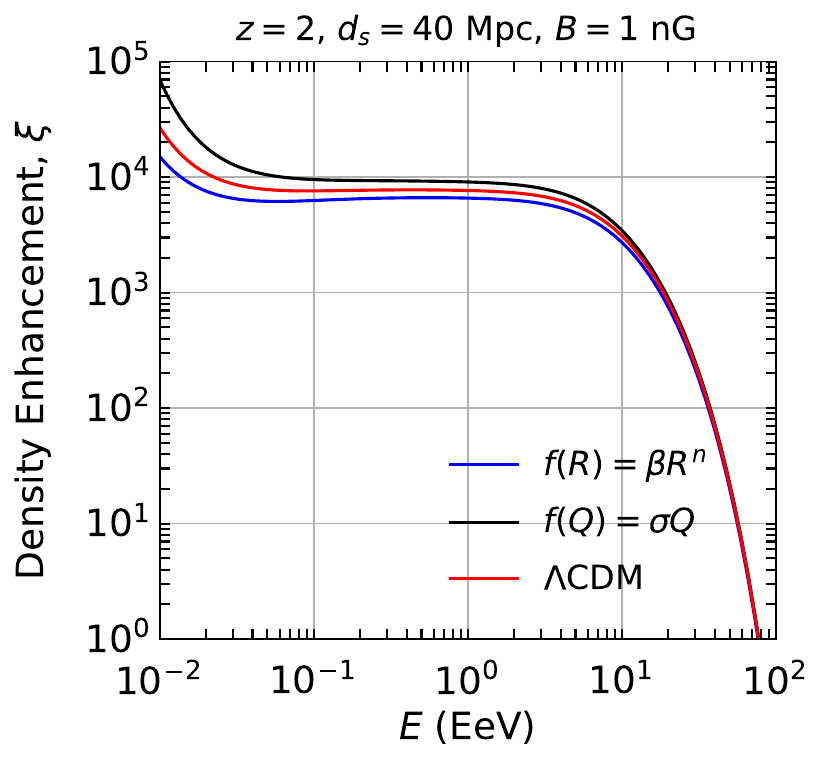}
\includegraphics[scale=0.4]{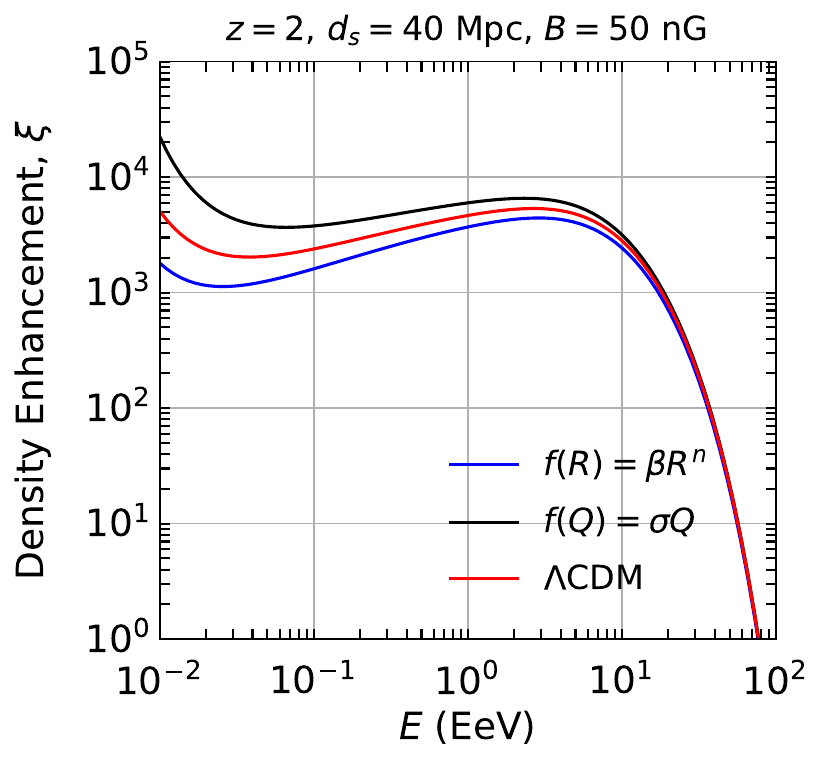}
\includegraphics[scale=0.4]{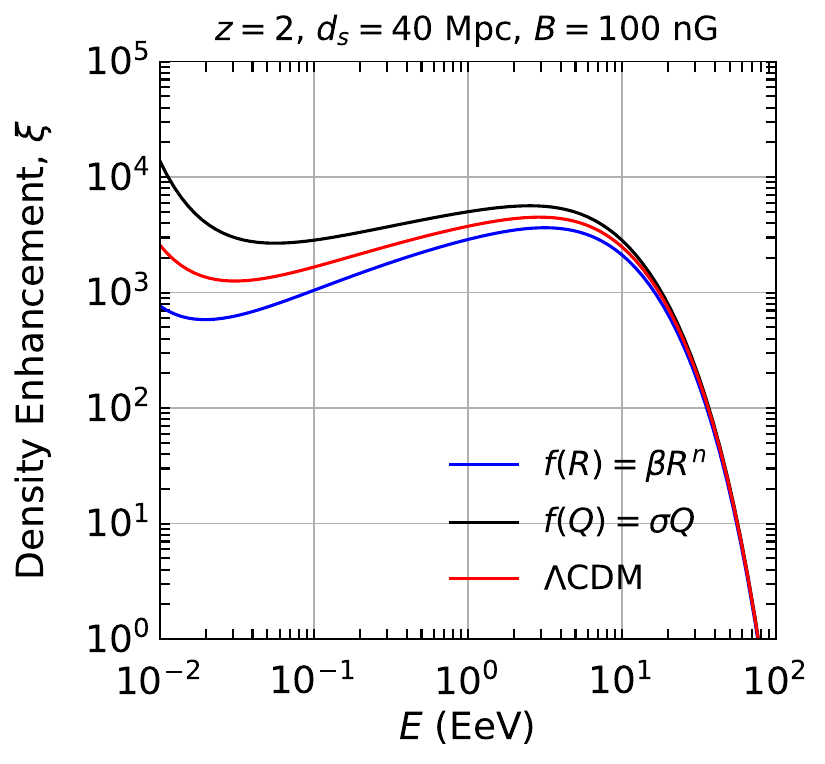}}
\vspace{-0.2cm}
\caption{The density enhancement factor $\xi$ as a function of energy $E$ for 
$z=2$ and $l_\text{c}=1$ Mpc by considering the magnetic field variations as 
predicted by the $f(R)$ gravity model, $f(Q)$ gravity model and $\Lambda$CDM 
model.}
\label{enhancement_comp2}
\end{figure}

\begin{figure}[htb!]
\centerline{
\includegraphics[scale=0.375]{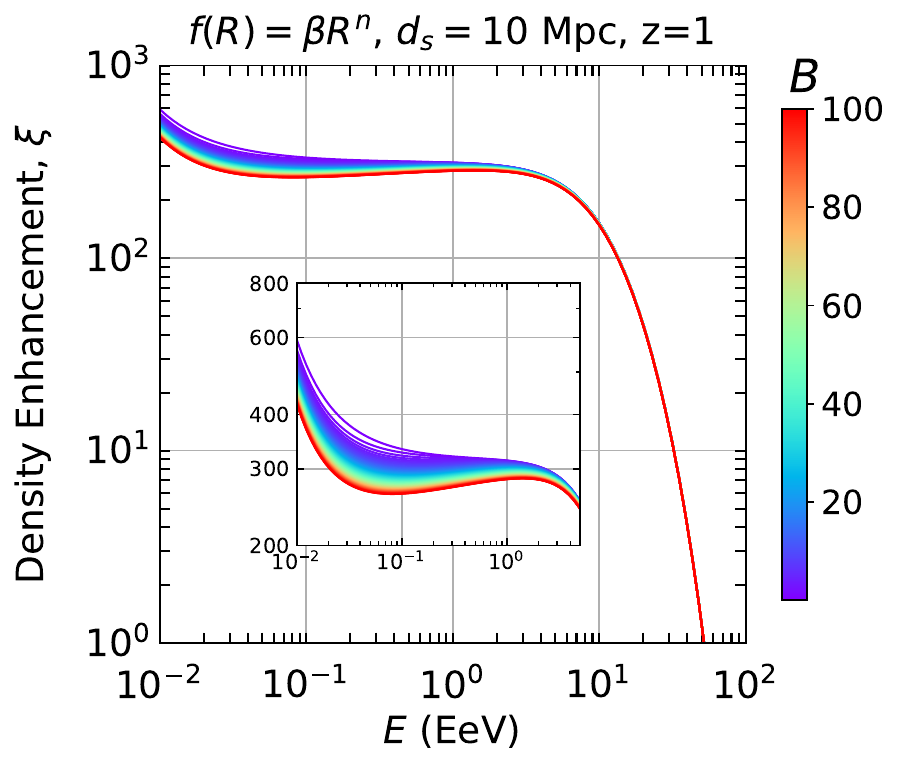}
\includegraphics[scale=0.375]{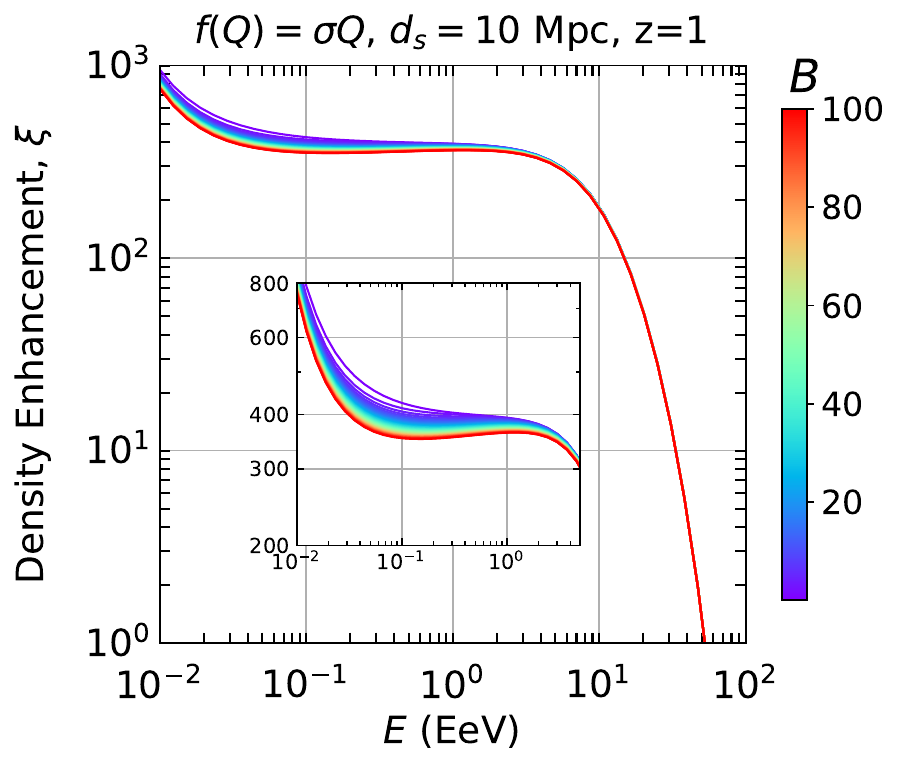}
\includegraphics[scale=0.375]{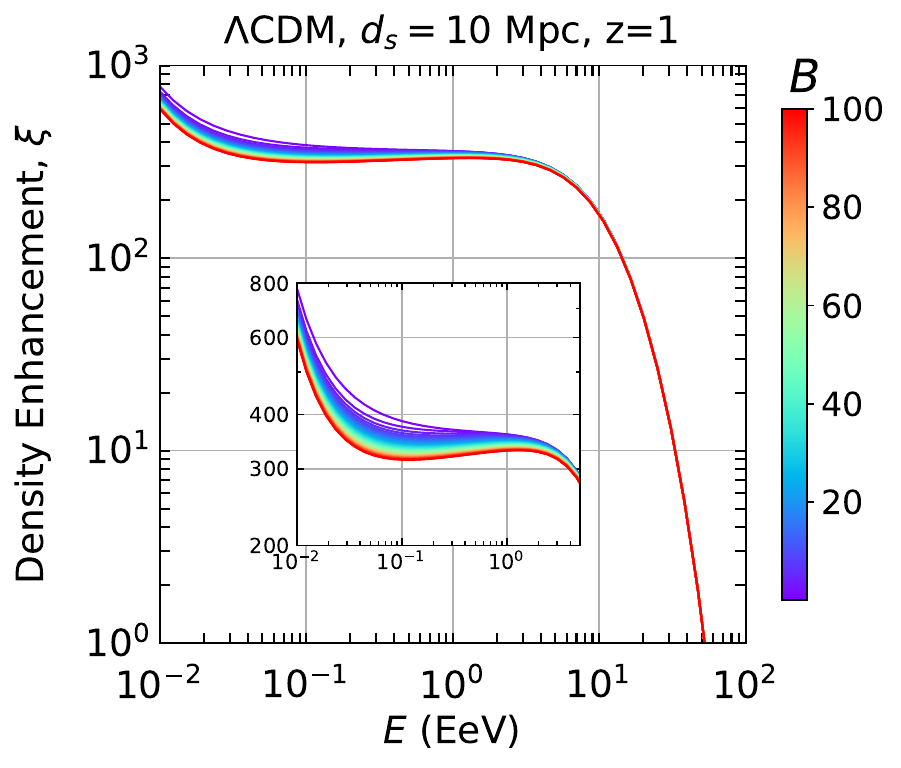}
}\vspace{0.2cm}
\centerline{
\includegraphics[scale=0.375]{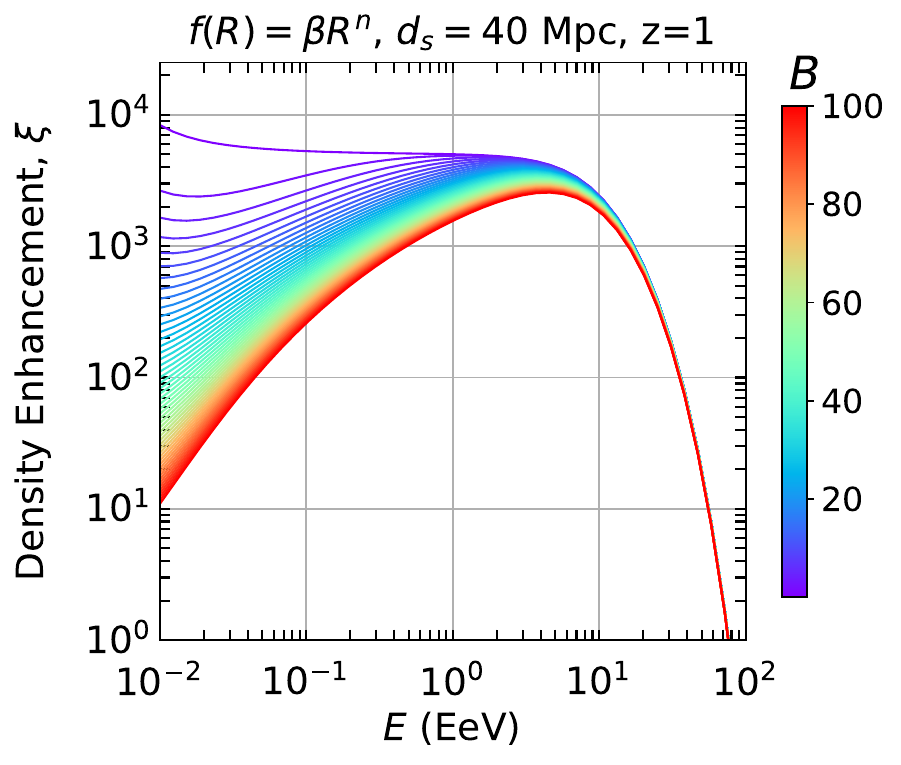}
\includegraphics[scale=0.375]{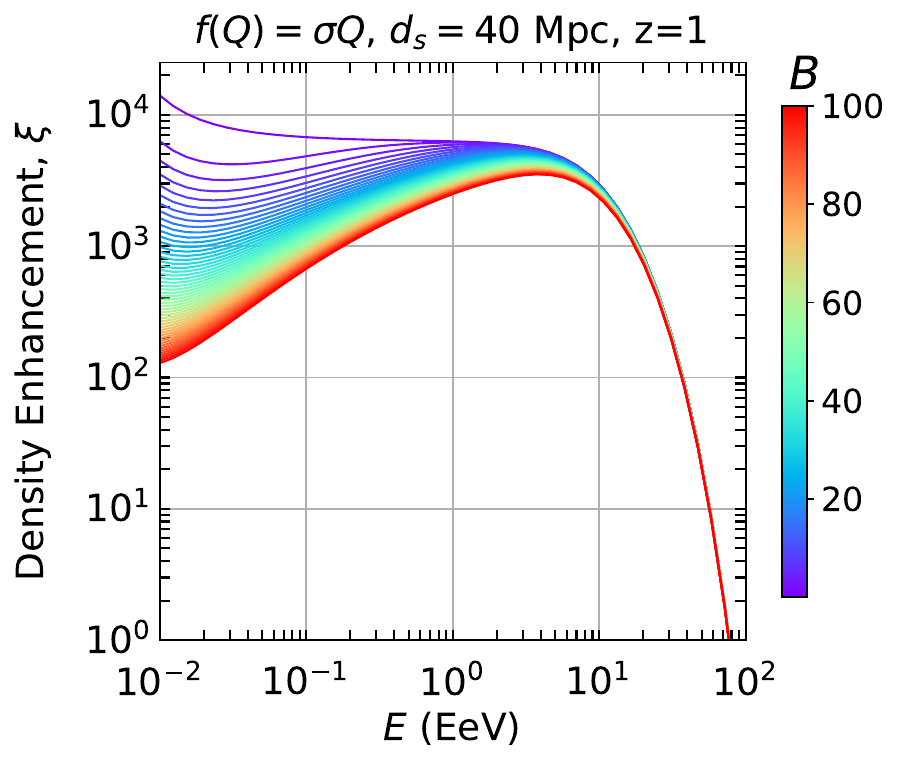}
\includegraphics[scale=0.375]{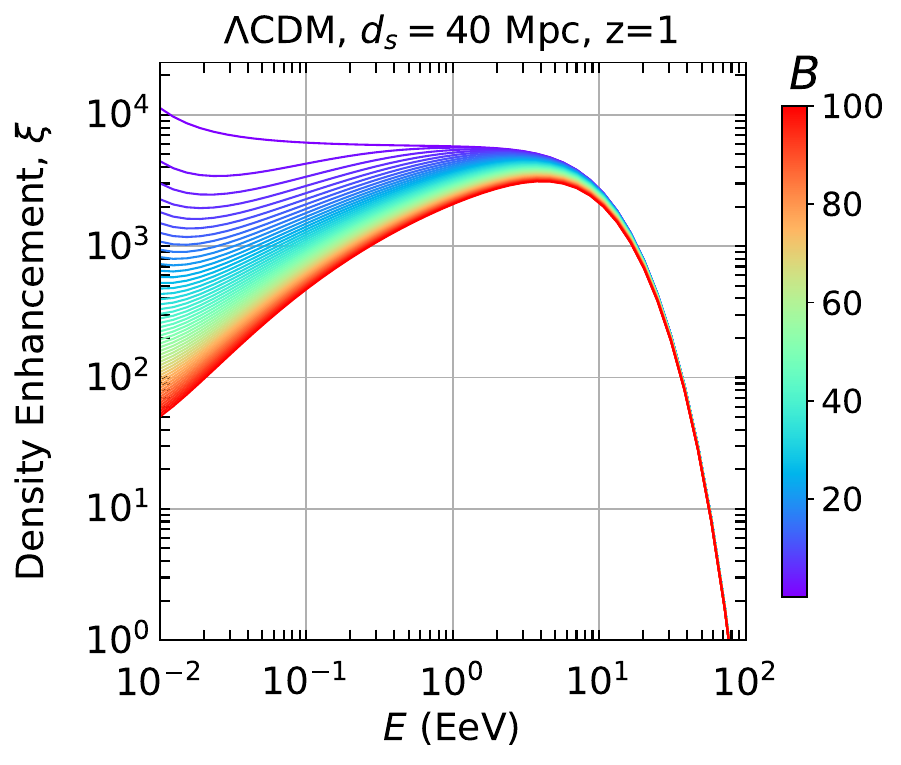}
}
\vspace{-0.2cm}
\caption{Top panel: The density enhancement factor $\xi$ as a function of 
energy $E$ for magnetic field strengths $B=1-100$ nG, $d_\text{s}=10$ Mpc, and 
$z=1$. The left, middle and right panels represent the $f(R)$ gravity 
model, $f(Q)$ gravity model and $\Lambda$CDM model respectively. Bottom panel: 
Same as the top panels but for $d_\text{s}=40$ Mpc.}
\label{fig_enhancement_b}
\end{figure}

Further comparative analysis between the cosmological models has been 
performed for the CR density enhancement by considering the variation of 
the magnetic field strength as $1$ nG, $50$ nG and $100$ nG as shown in 
Fig.~\ref{enhancement_comp2}. One can see that as the field strength 
increases, the enhancement factor tends to decrease along with increasing 
differences between models' predictions. The difference between models is
more pronounced in the lower energy side and the effects of models are 
appreciable only up to $10$ EeV. We also draw Fig.~\ref{fig_enhancement_b} to 
give a clearer justification of the magnetic field range and its effects.
In Fig.~\ref{fig_enhancement_b}, we plot the enhancement factor with 
respect to the energy for the different magnetic field amplitudes. We set the 
redshift value  $z=1$, source separation distance $d_\text{s}=10$ Mpc, and 
$l_\text{c}=1$ Mpc for this calculation. The amplitude of the magnetic field 
varies from $1$ to $100$ nG. The left, middle, and right panels represent the 
$f(R)$ model, $f(Q)$ model, and the $\Lambda$CDM model, respectively.
An inset plot is drawn in each of the cosmological model cases. From the inset 
plot, it is visible that the cosmological model has a significant effect on 
the CRs propagation. In both high and low energy regions the $f(Q)$  model 
predicts the highest enhancement then followed by the $\Lambda$CDM and the 
$f(R)$ model, as mentioned earlier. In the low energy region, the effect of the 
cosmological models is evident in all ranges of the magnetic field strength 
considered here. However, the cosmological models' effect is less in the 
higher energy range since the magnetic field does not have any significant 
contribution to the enhancement of CRs within this range of energy. The bottom 
panel of Fig.~\ref{fig_enhancement_b} shows the same situation but at the 
distance between the sources $d_\text{s}=40$ Mpc. In this case, the magnetic 
field can significantly contribute to the enhancement factor up to $10$ EeV, 
after that, all contributions from the magnetic field seem to be the same. In 
the $f(R)$ model, the effect of the magnetic field is comparatively higher as 
it covers the enhancement amplitude from about $10$ to  $9000$. However, the 
$f(Q)$ model predicts the highest enhancement as expected. Thus, depending on 
the redshift, magnetic field, distance between the sources, and the energy, the 
cosmological models have significant effects on the propagation of the CRs. 

\begin{figure}[t!]
\centerline{
\includegraphics[scale=0.4]{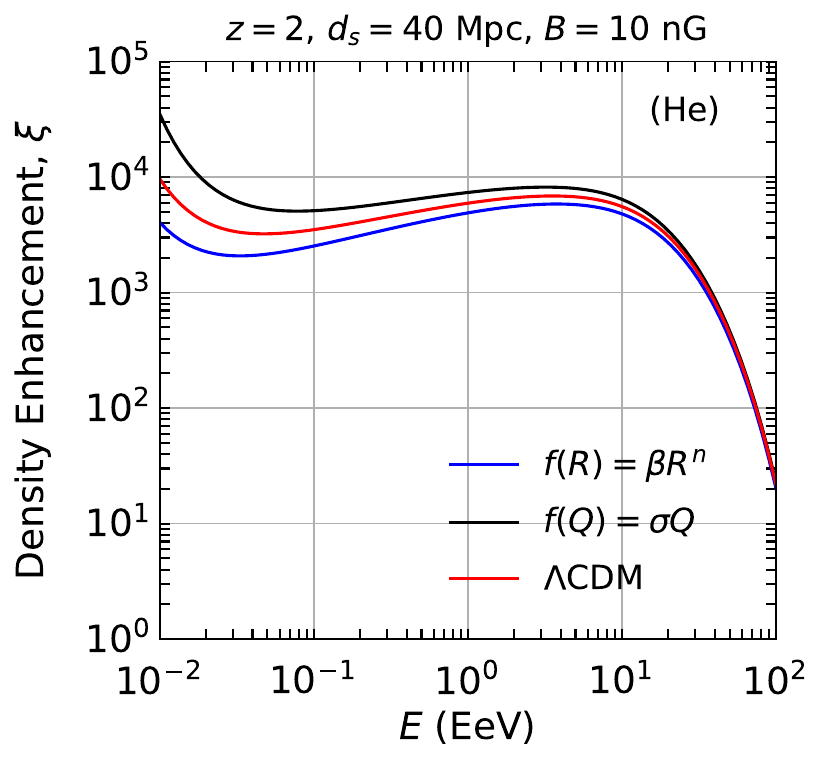}
\includegraphics[scale=0.4]{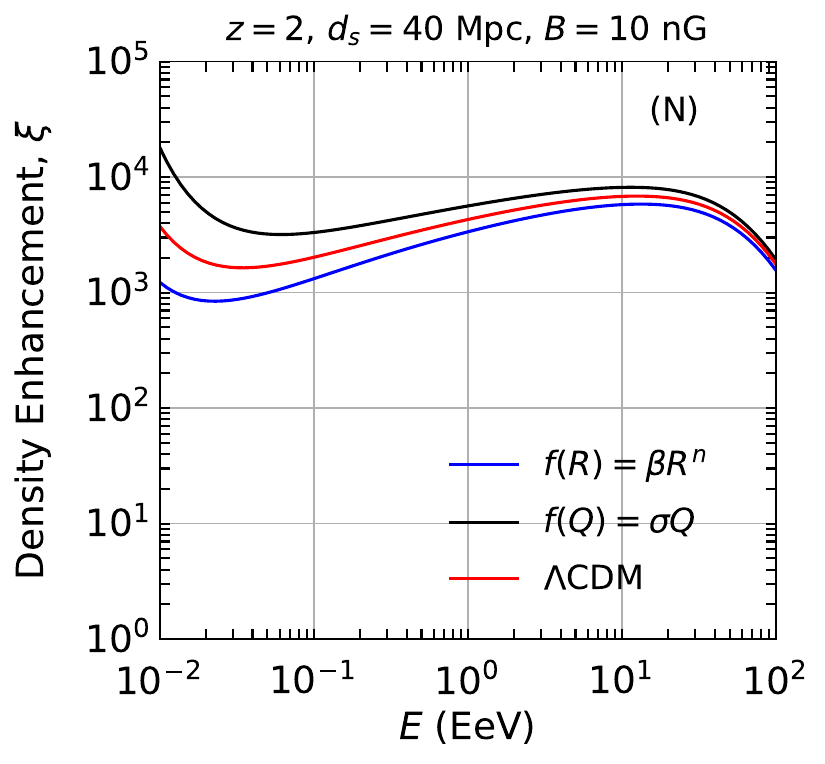}
\includegraphics[scale=0.4]{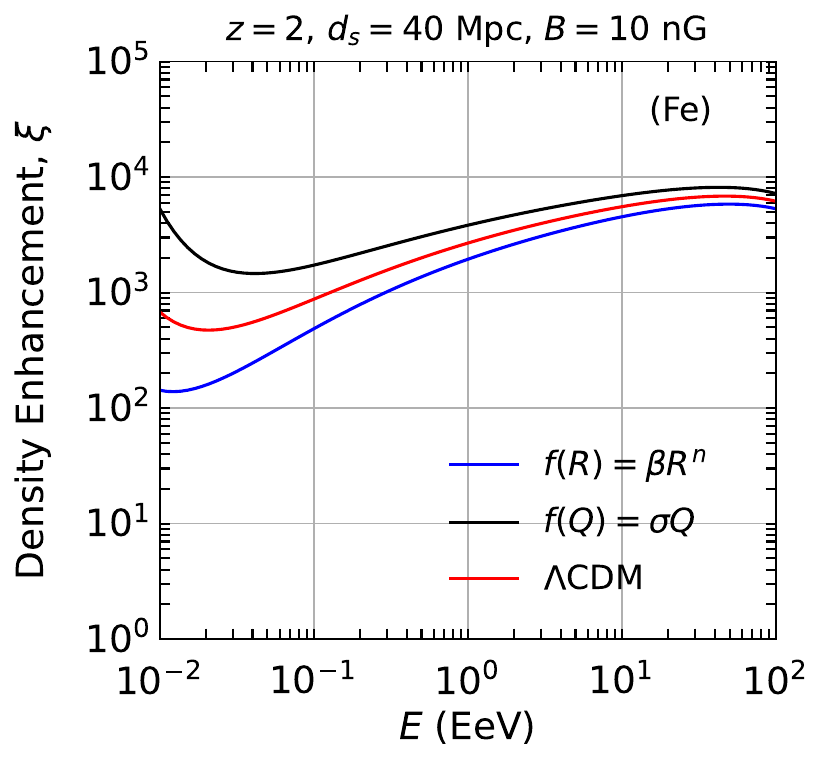}}
\vspace{-0.2cm}
\caption{The density enhancement factor $\xi$ as a function of energy $E$ with 
$z=2$, $B=10$ nG and $l_\text{c}=1$ Mpc as predicted by the $f(R)$ gravity
model, $f(Q)$ gravity model and $\Lambda$CDM model for He (left panel), N 
(middle panel), and Fe (right panel) nuclei.}
\label{enhancement_comp3}
\end{figure}

\begin{figure}[h]
\centerline{
\includegraphics[scale=0.375]{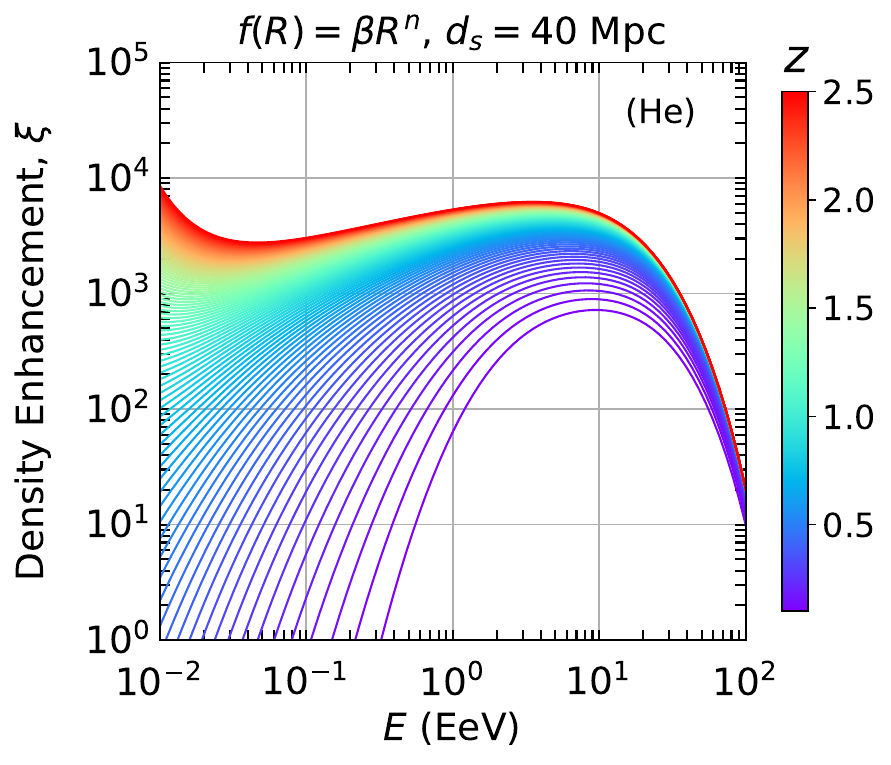}
\includegraphics[scale=0.375]{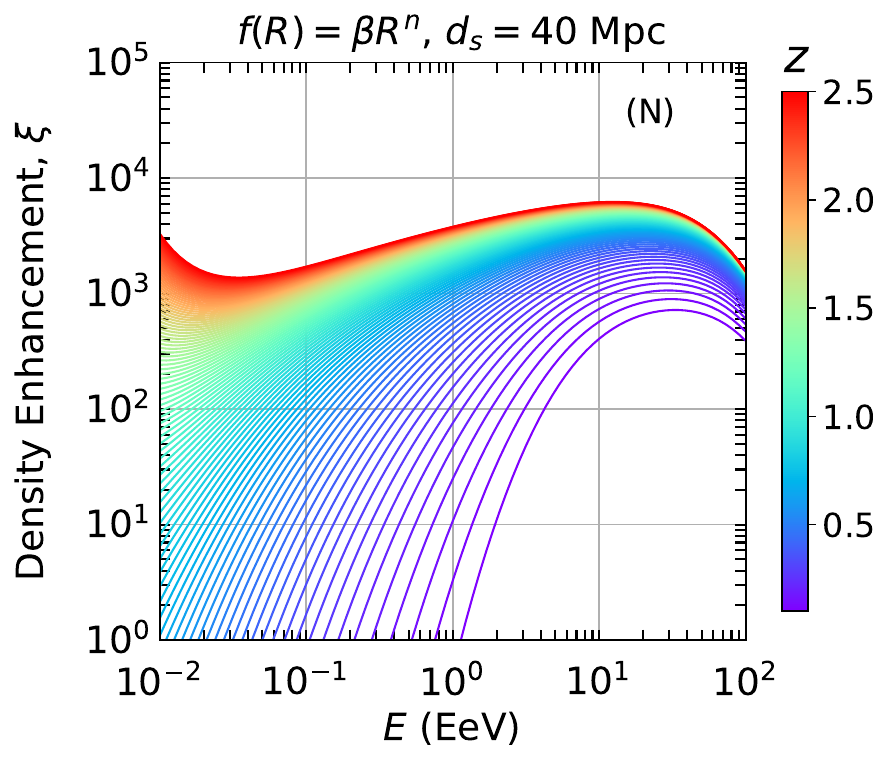}
\includegraphics[scale=0.375]{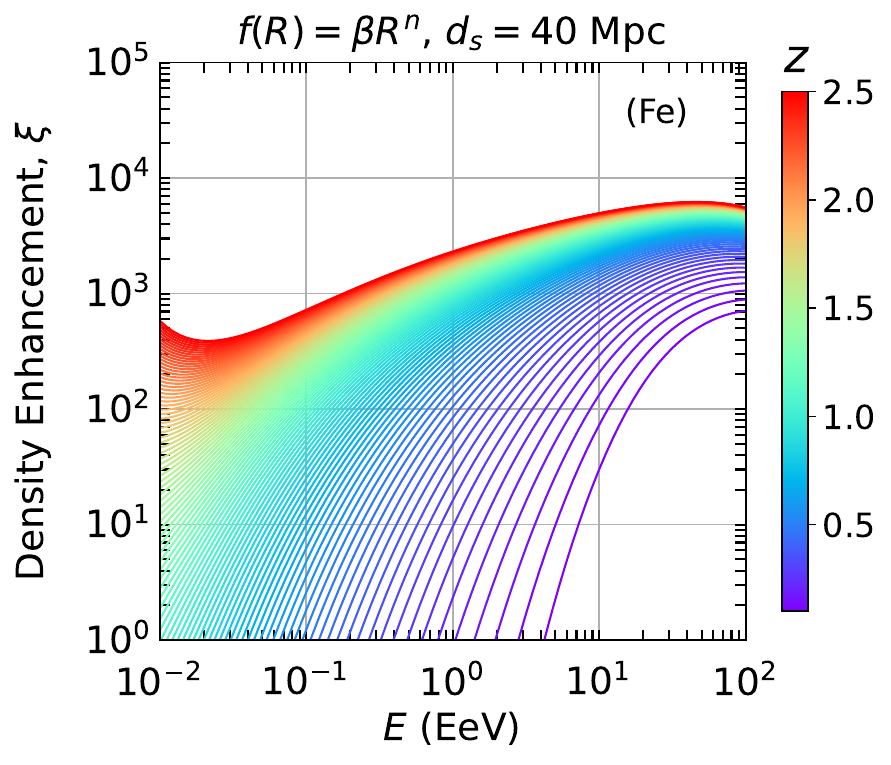}
}\vspace{0.2cm}
\centerline{
\includegraphics[scale=0.375]{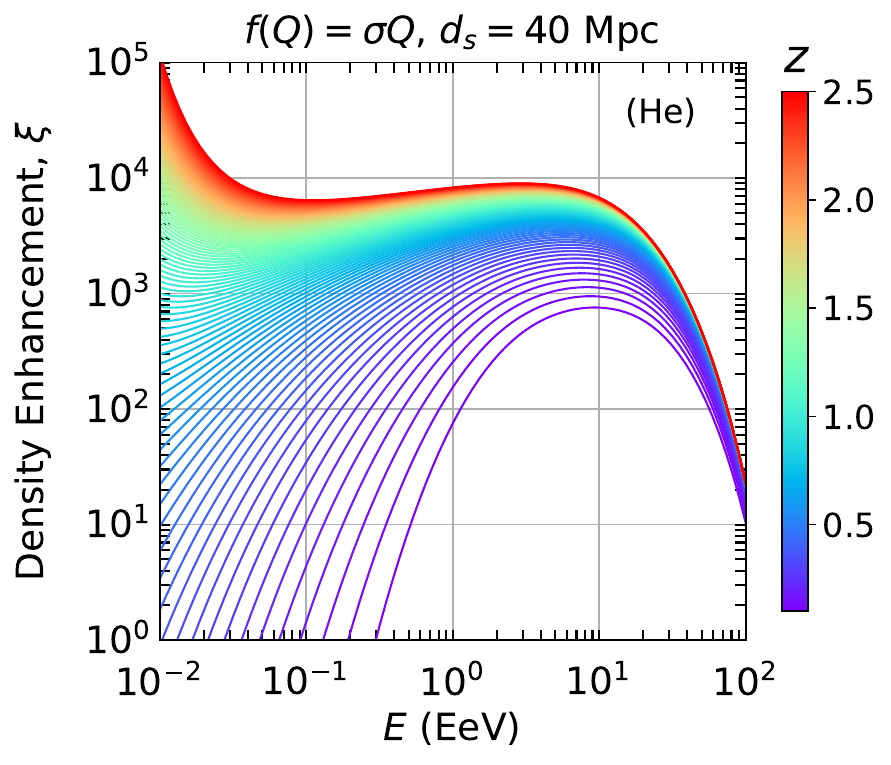}
\includegraphics[scale=0.375]{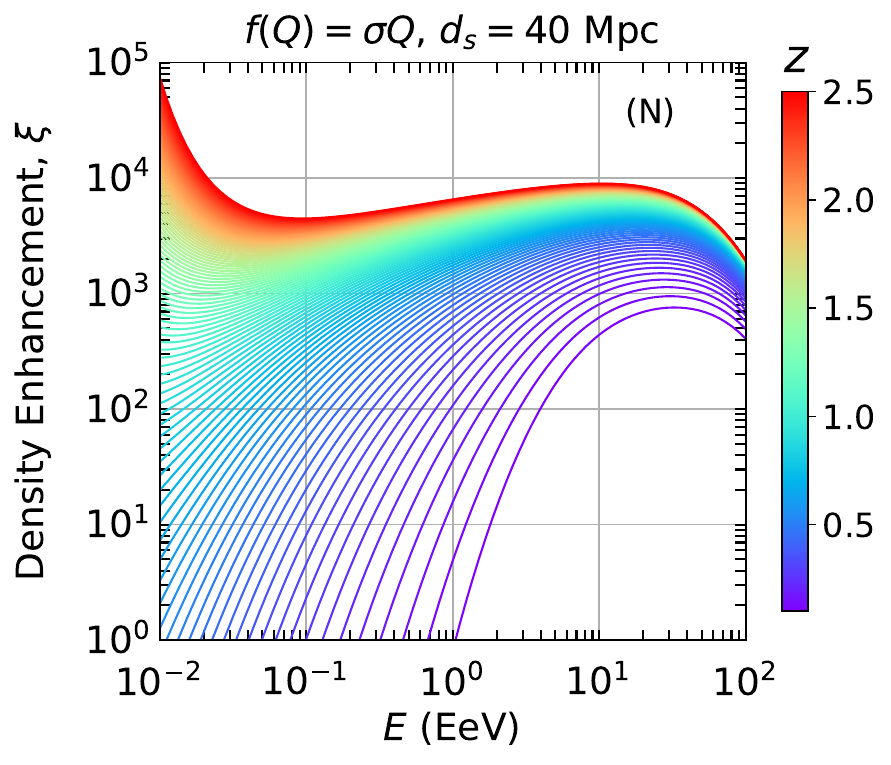}
\includegraphics[scale=0.375]{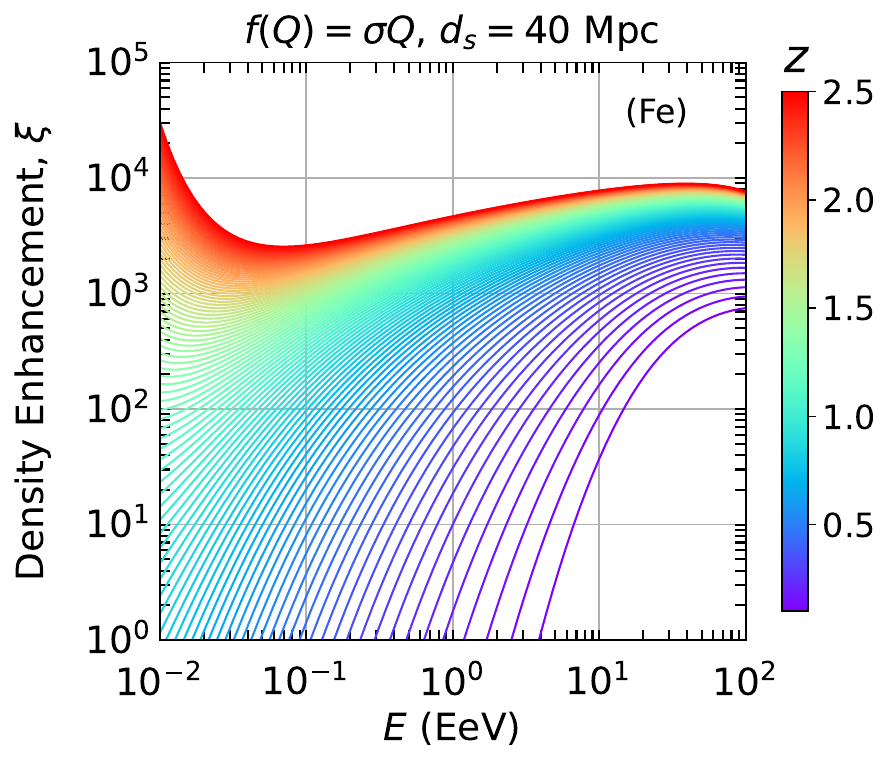}
}\vspace{0.2cm}
\centerline{
\includegraphics[scale=0.375]{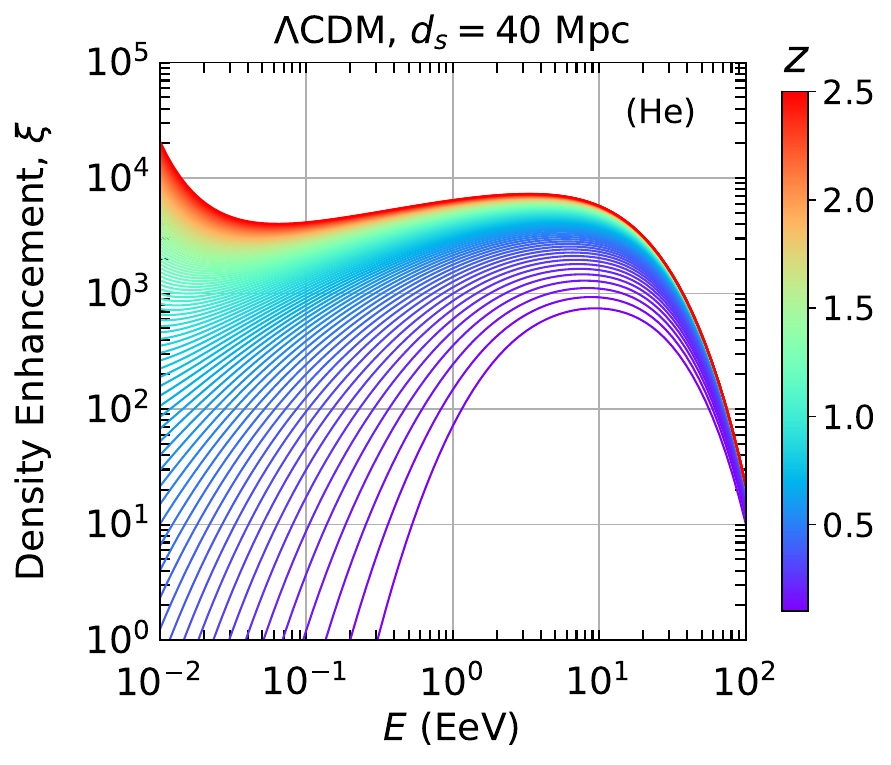}
\includegraphics[scale=0.375]{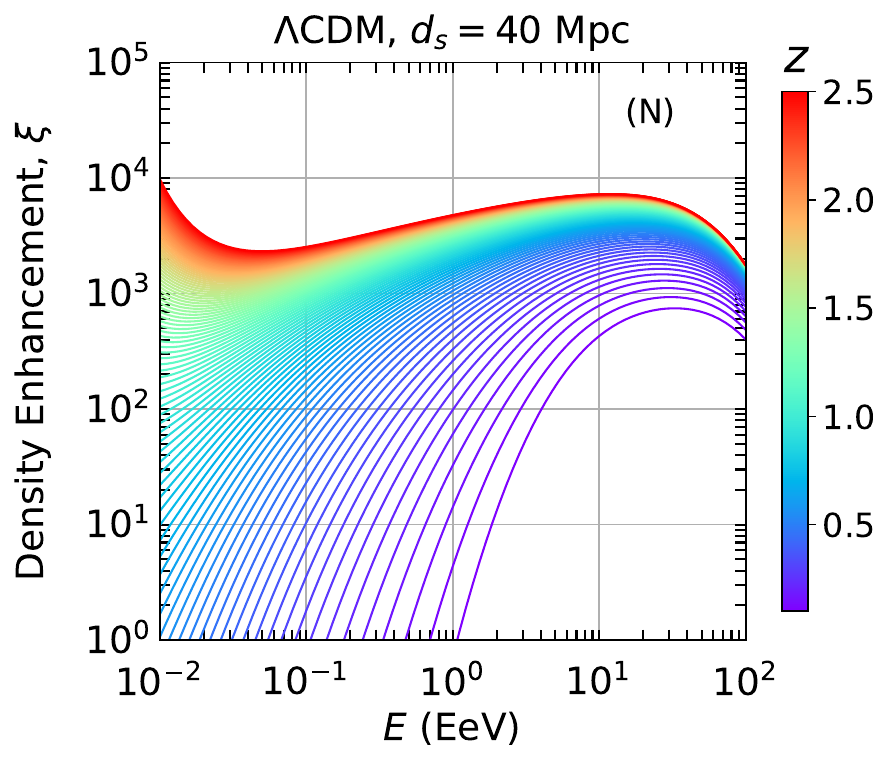}
\includegraphics[scale=0.375]{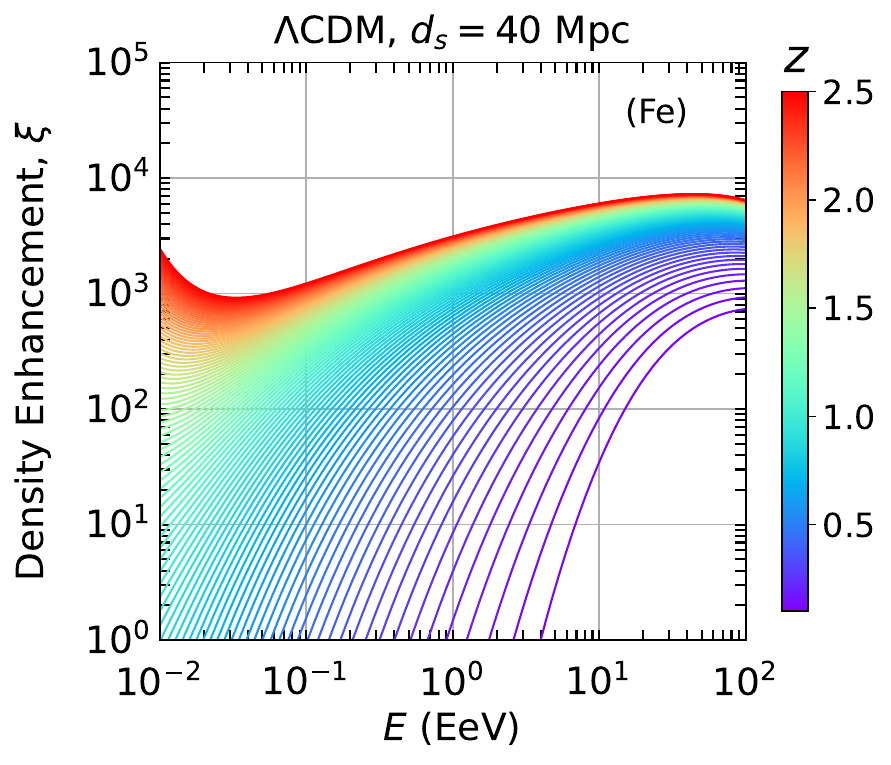}
}
\vspace{-0.2cm}
\caption{The density enhancement factor $\xi$ for different nuclei as a 
function of energy $E$ for $z=0-2.5$ by considering $B=10$ nG, $l_\text{c}=1$ 
Mpc. The top panels represent the results from the $f(R)$ gravity model, the 
middle panels represent the results from the $f(Q)$ gravity model, and the 
bottom panels represent the results from the standard $\Lambda$CDM model. The 
considered nuclei are He, N, and Fe (left to right).}
\label{fig_enhancement_nuclei}
\end{figure}

For the nuclei (He, N, and Fe), we simply perform a comparative analysis with 
the results from the $f(R)$, $f(Q)$ and $\Lambda$CDM models for the redshift 
$z=2$ as shown in Fig.~\ref{enhancement_comp3}. For the heavy nuclei, 
the cosmological models' effect is more visible in both low and high energy 
ranges. Moreover, for a better understanding of the redshift effect, we plot 
Fig.~\ref{fig_enhancement_nuclei}. For the He nuclei, the enhancement results 
for the $f(Q)$ gravity model predict the highest one for the energy range 
between $0.01-10$ EeV. For $E > 10$ EeV, all these cosmological models predict 
the same kind of results. The same results can also be seen in the CNO group. 
But in the case of Fe nuclei, we see some interesting and different kinds of 
results. For these nuclei, the density enhancement parameter is gradually 
increasing after the energy of about $0.02$ EeV, $0.05$ EeV and $0.035$ EeV 
for the $f(R)$ power-law model, $f(Q)$ model and $\Lambda$CDM model 
respectively. It needs to be mentioned that these results are somewhat 
different from the results of previous literature \cite{mollerech2019, swaraj1}
due to the reason that in our work we consider the discrete source 
distribution as mentioned in Eq.~\eqref{ri} and the considered redshift 
range is also different in the mentioned literature.

\begin{figure}[htb!]
\centerline{
\includegraphics[scale=0.36]{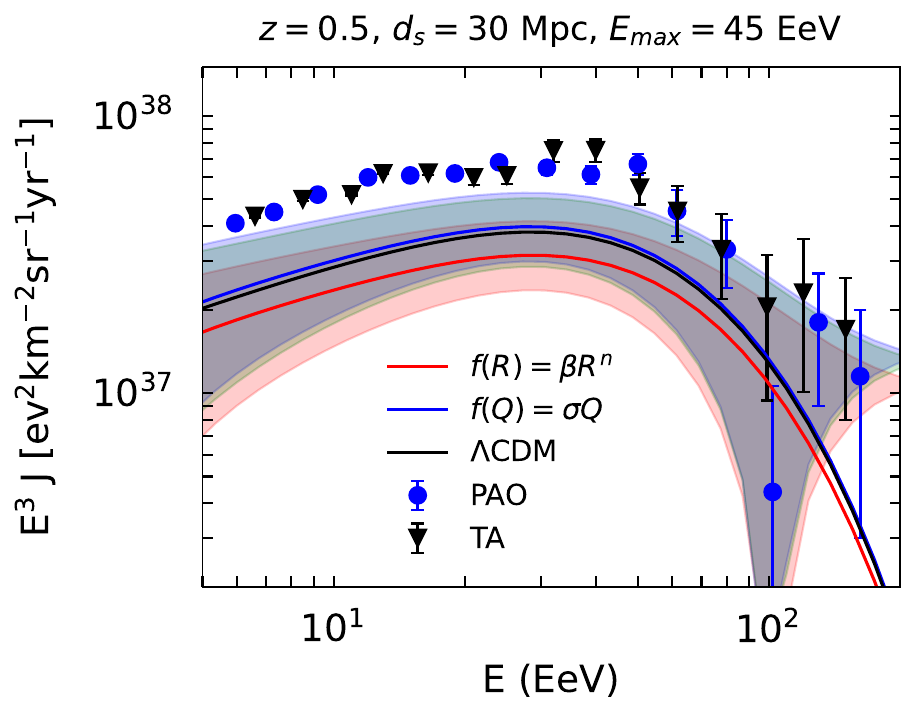}
\includegraphics[scale=0.36]{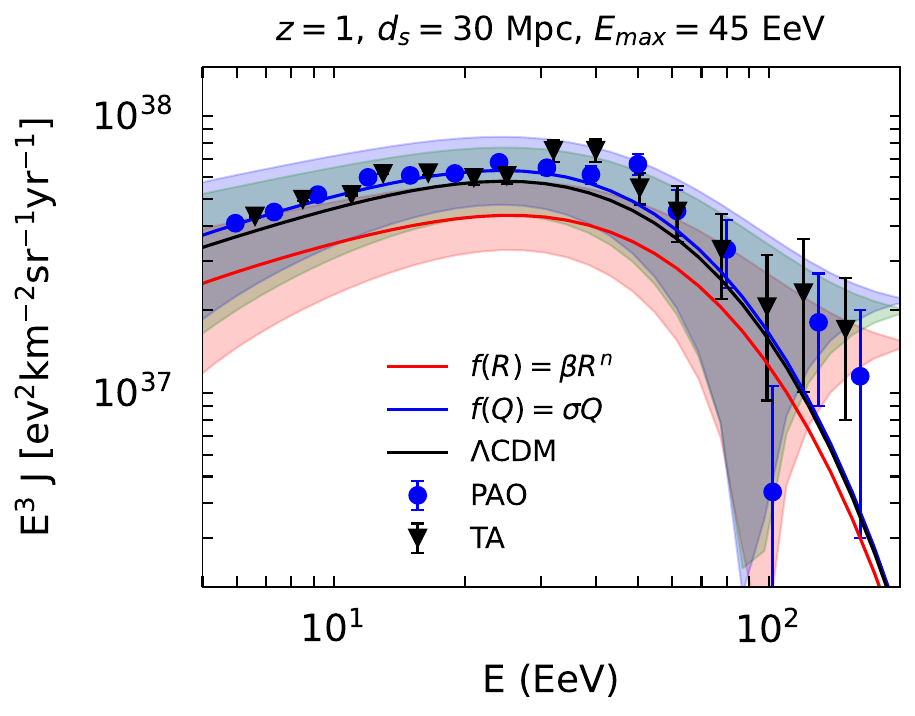}
\includegraphics[scale=0.36]{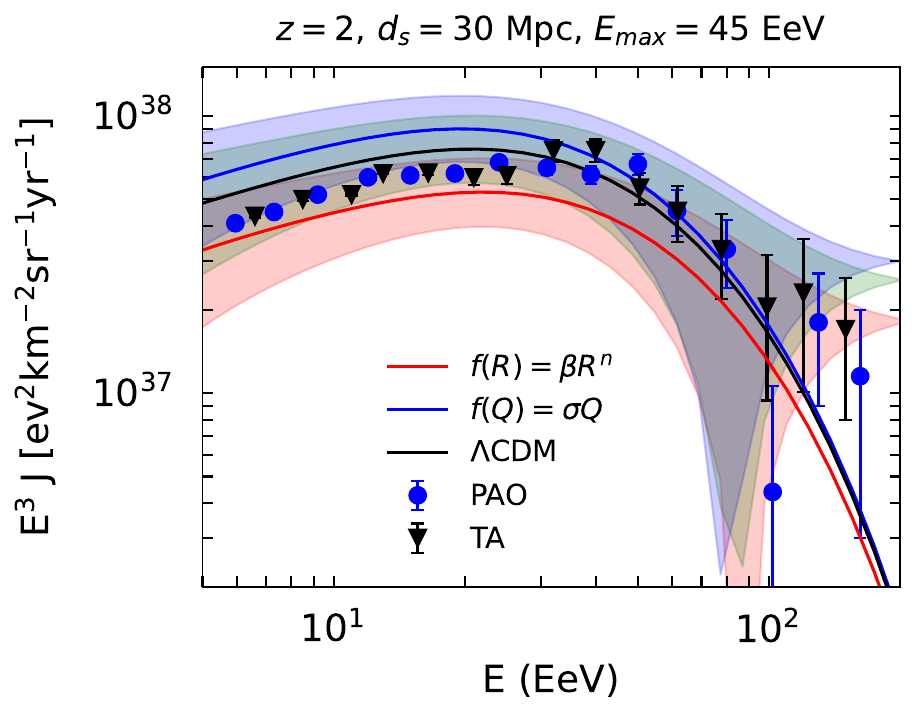}}
\centerline{
\includegraphics[scale=0.36]{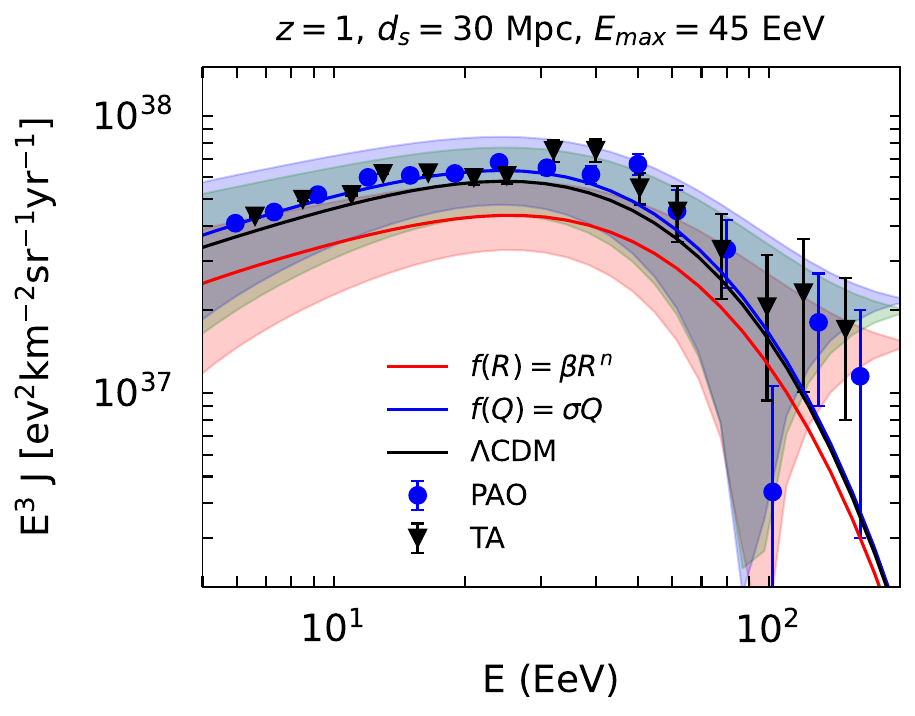}
\includegraphics[scale=0.36]{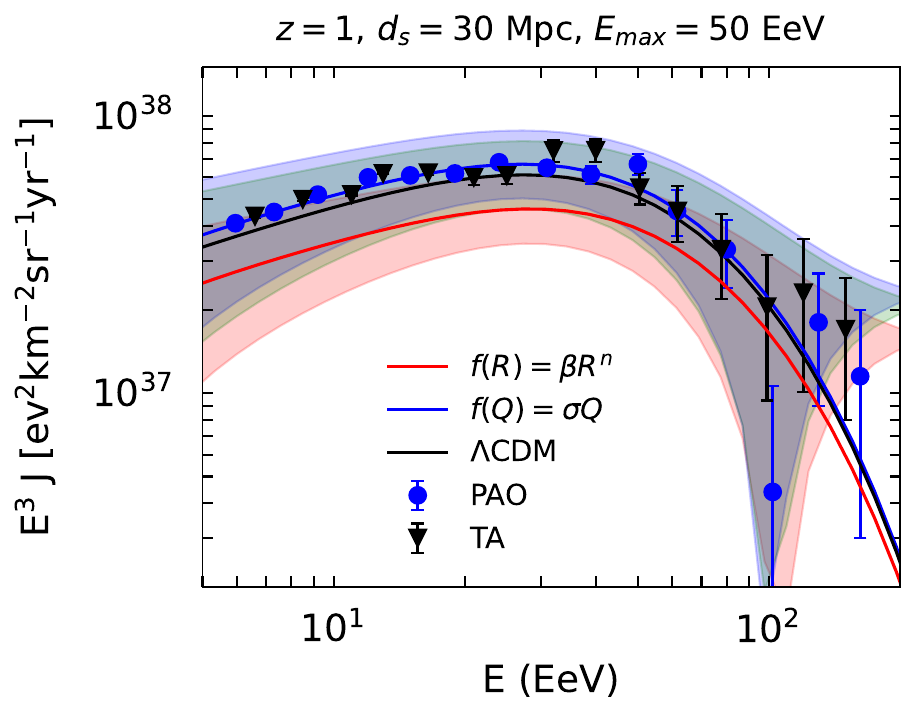}
\includegraphics[scale=0.36]{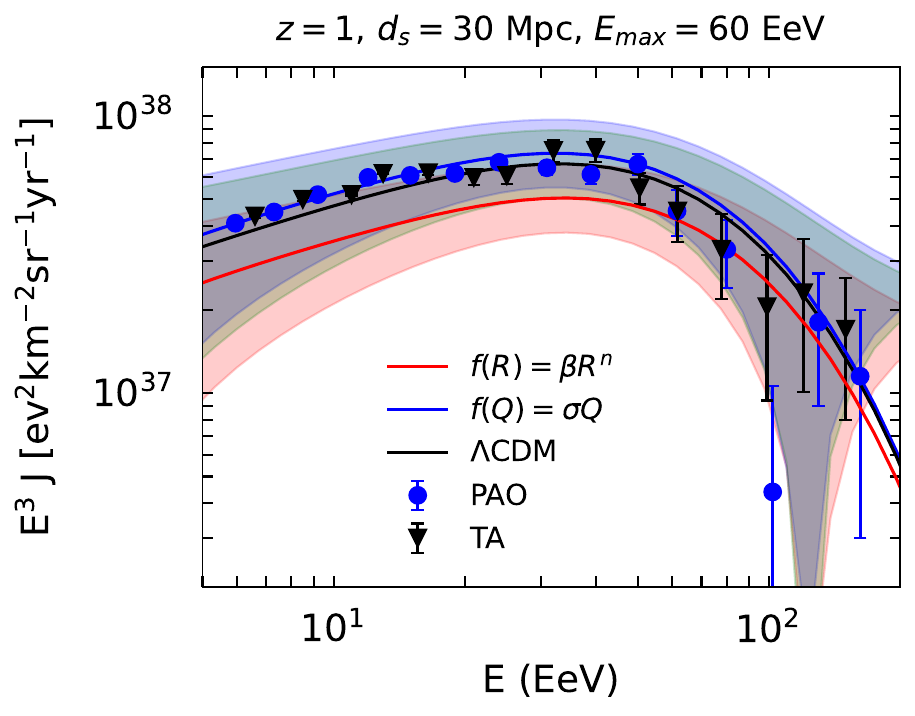}}
\vspace{-0.2cm}
\caption{UHECRs flux comparison for the $f(R)$ gravity model, $f(Q)$ gravity 
model, and $\Lambda$CDM model as obtained for different scenarios (see text). 
The observational data are from PAO \cite{augerprd2020} and TA \cite{ta2019}.
The shaded regions depict the uncertainty in the flux predicted by the different cosmological models.}
\label{flux_comp}
\end{figure} 

\begin{figure}
\centerline{
\includegraphics[scale=0.28]{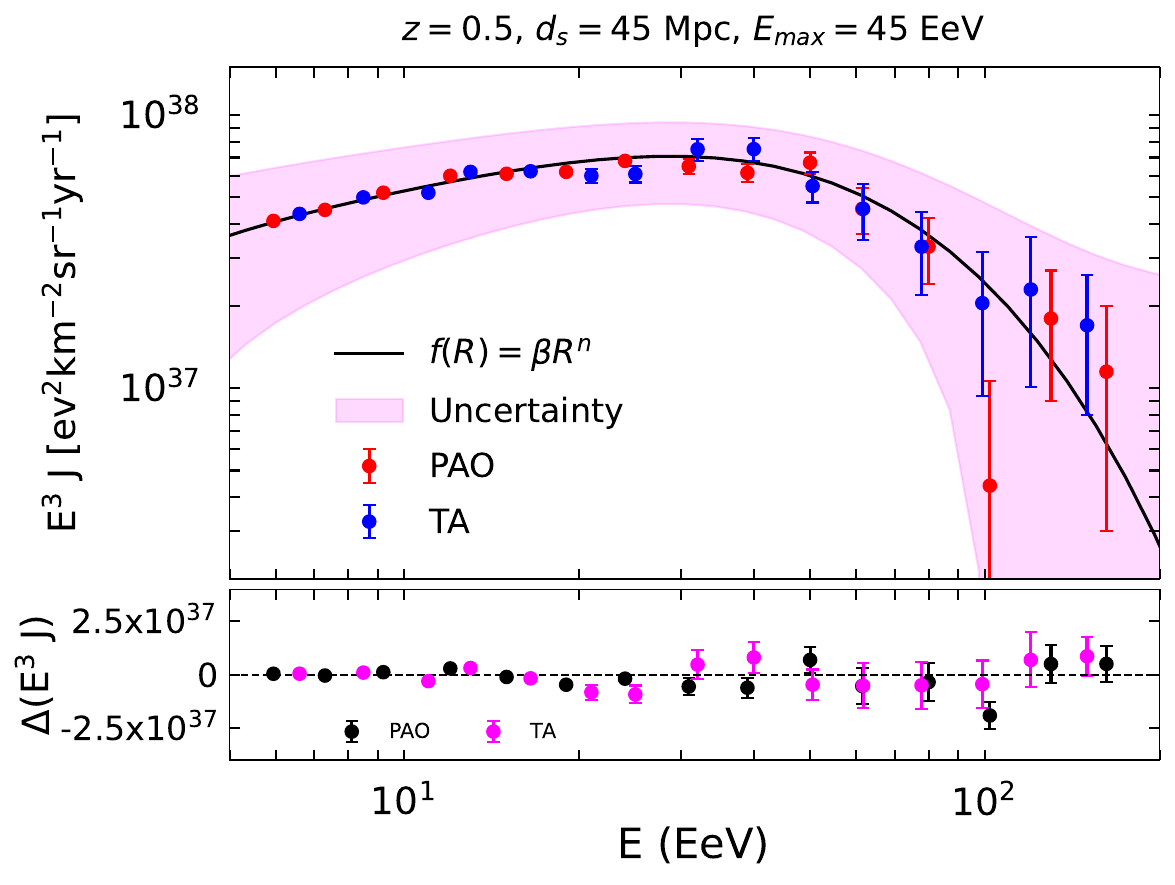}
\includegraphics[scale=0.28]{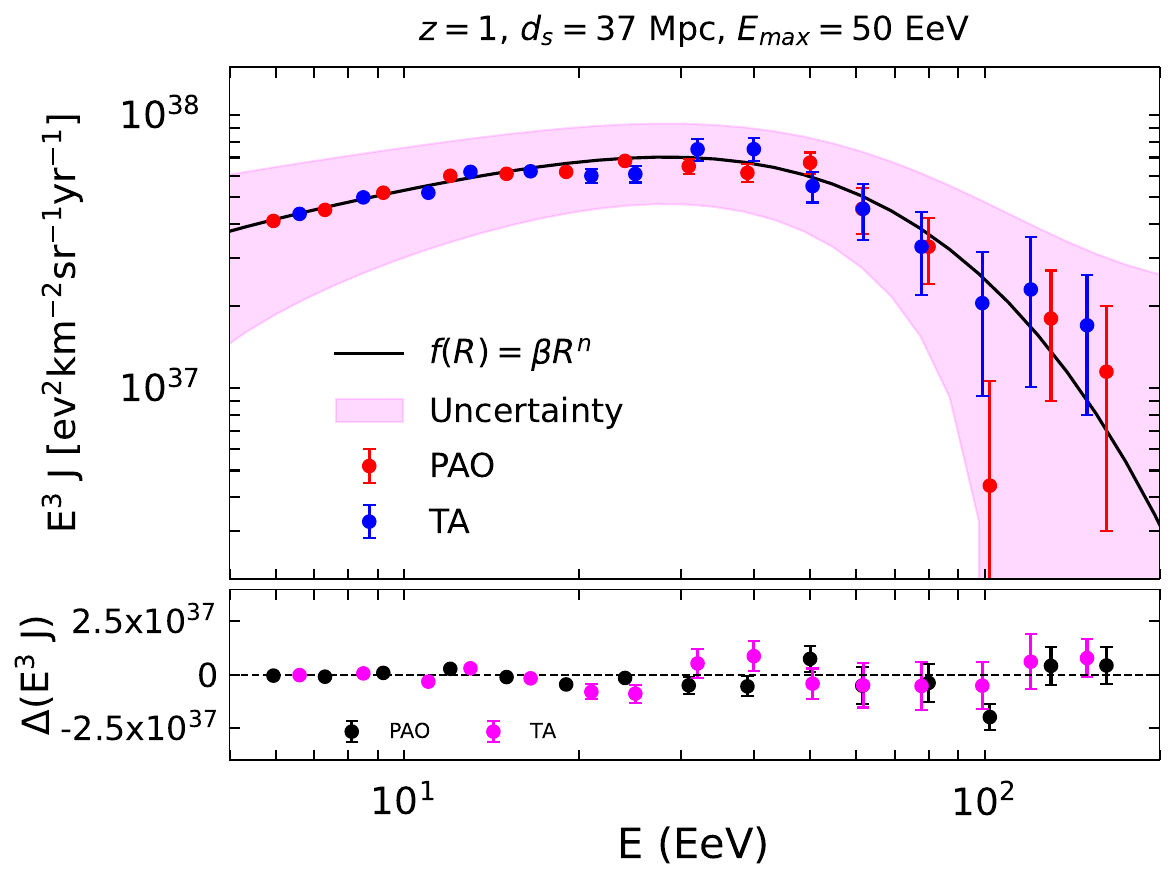}
\includegraphics[scale=0.28]{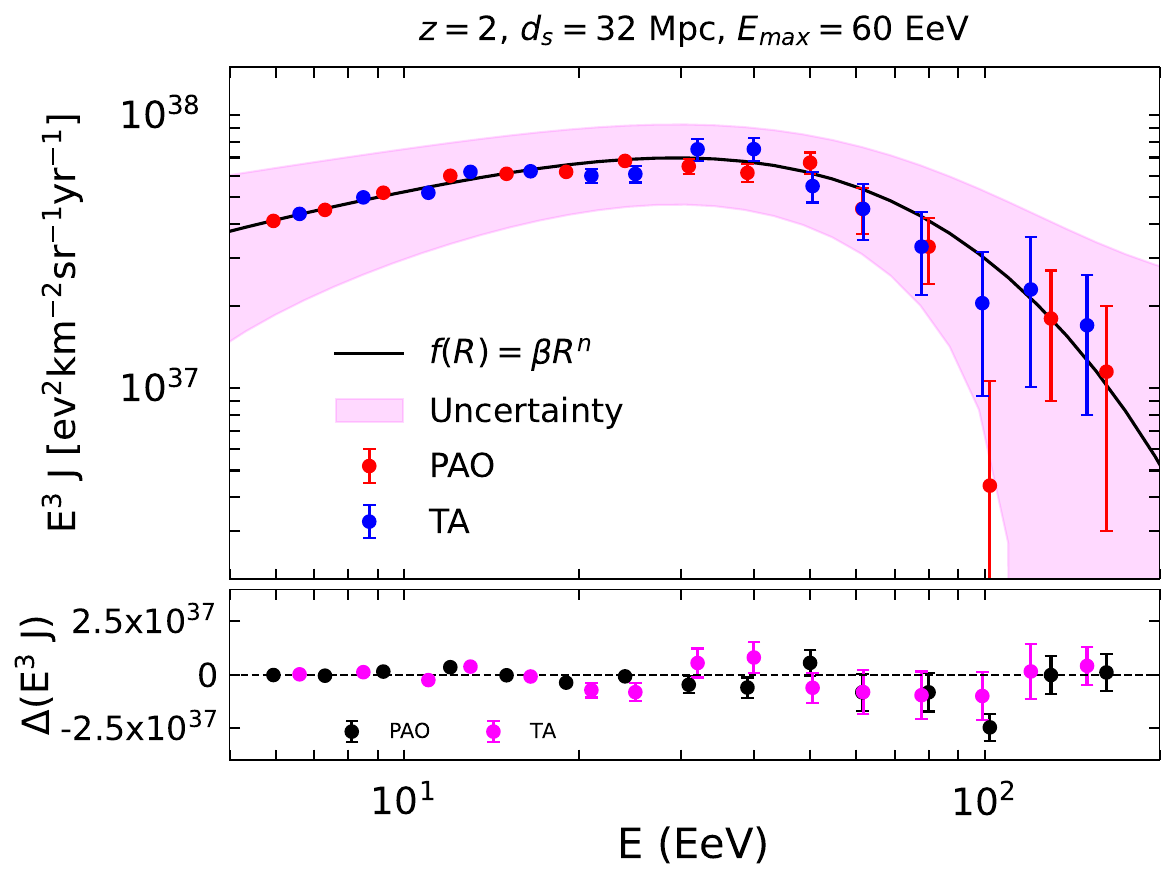}}\vspace{0.2cm}
\centerline{
\includegraphics[scale=0.28]{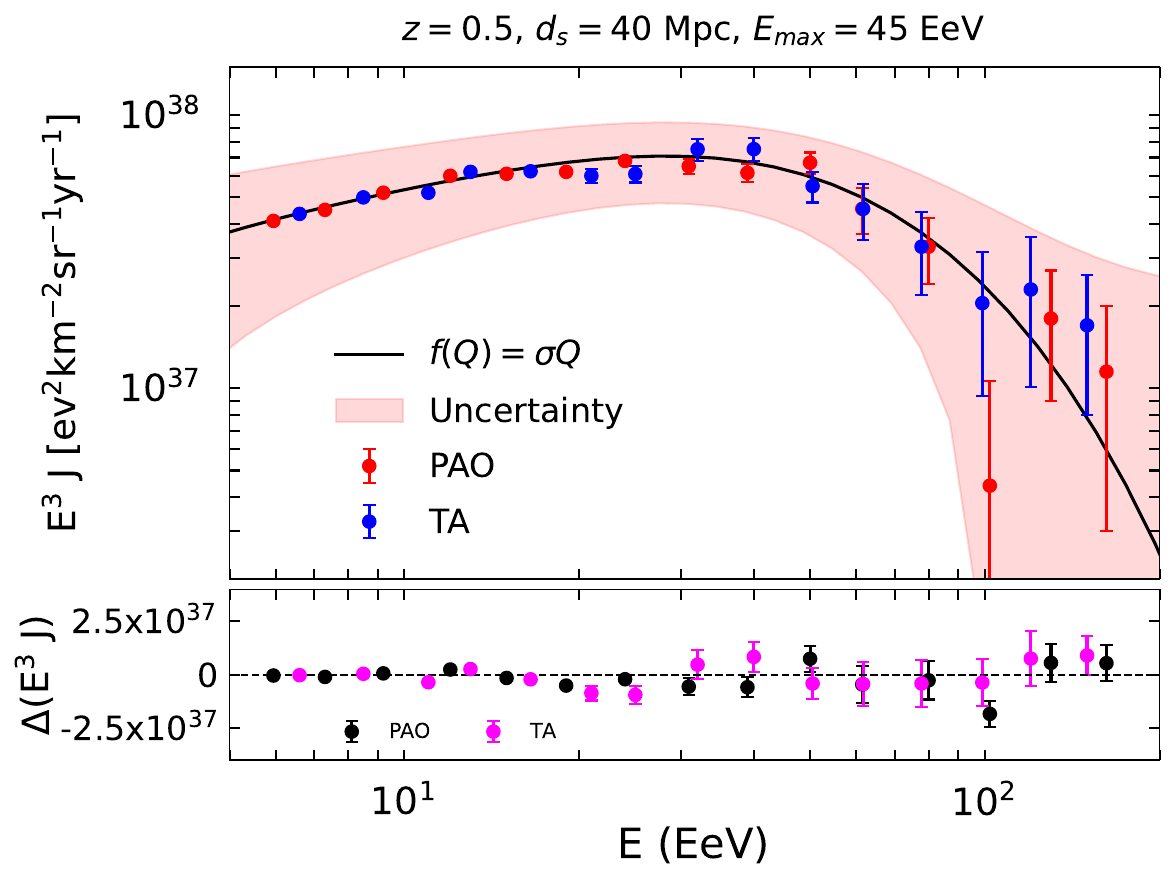}
\includegraphics[scale=0.28]{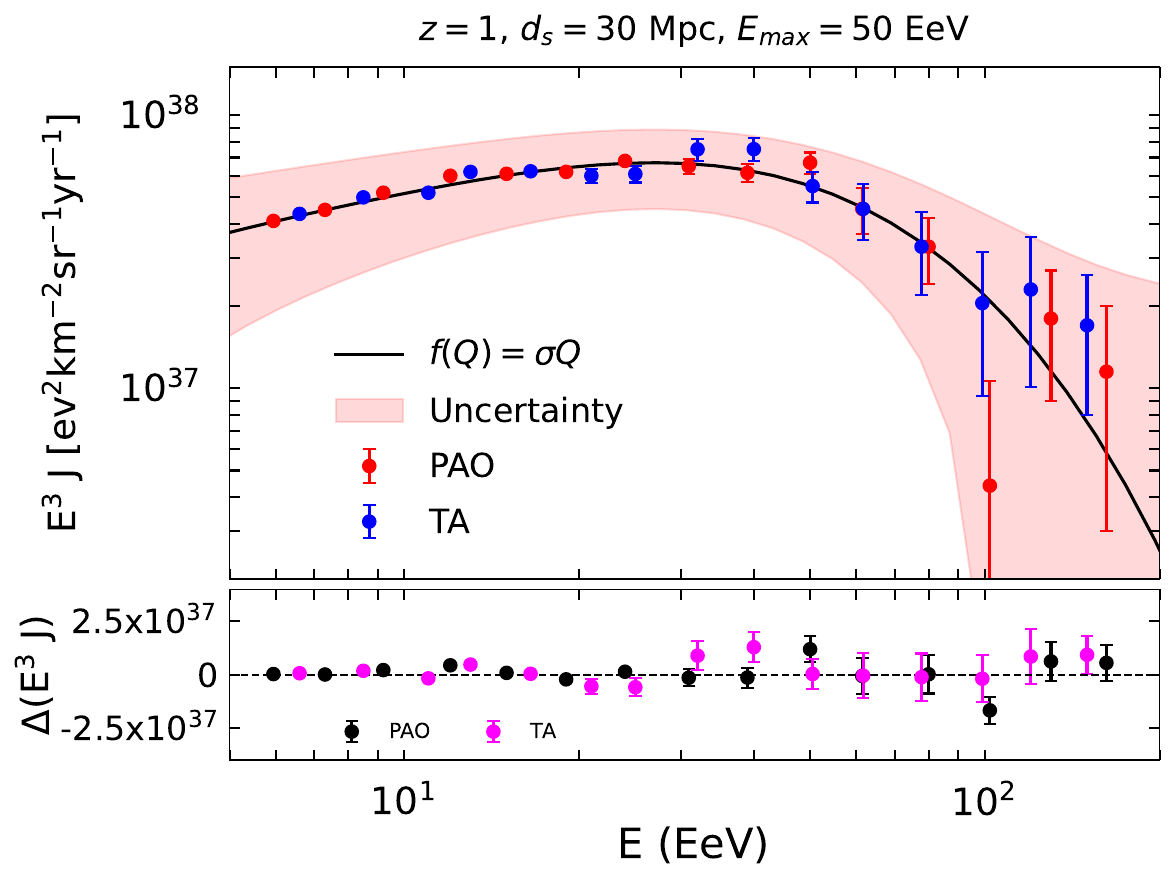}
\includegraphics[scale=0.28]{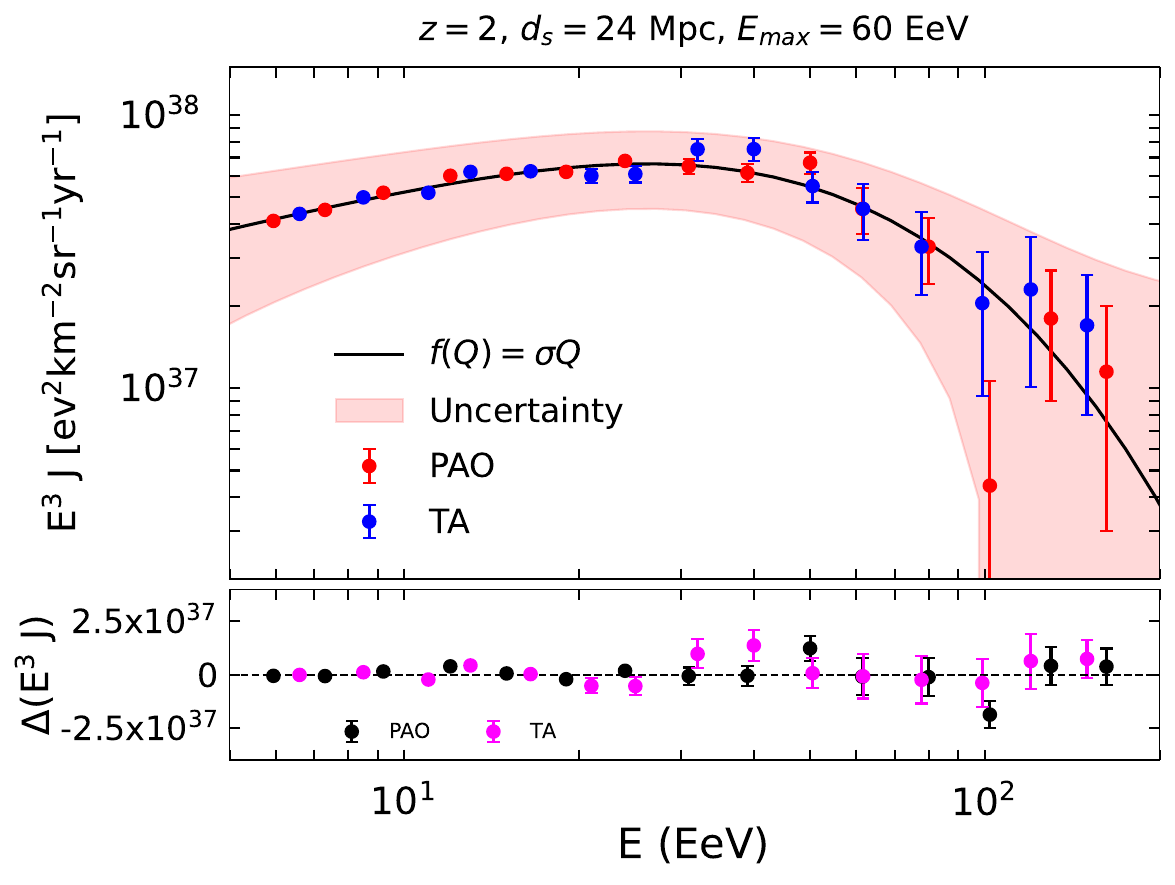}}\vspace{0.2cm}
\centerline{
\includegraphics[scale=0.28]{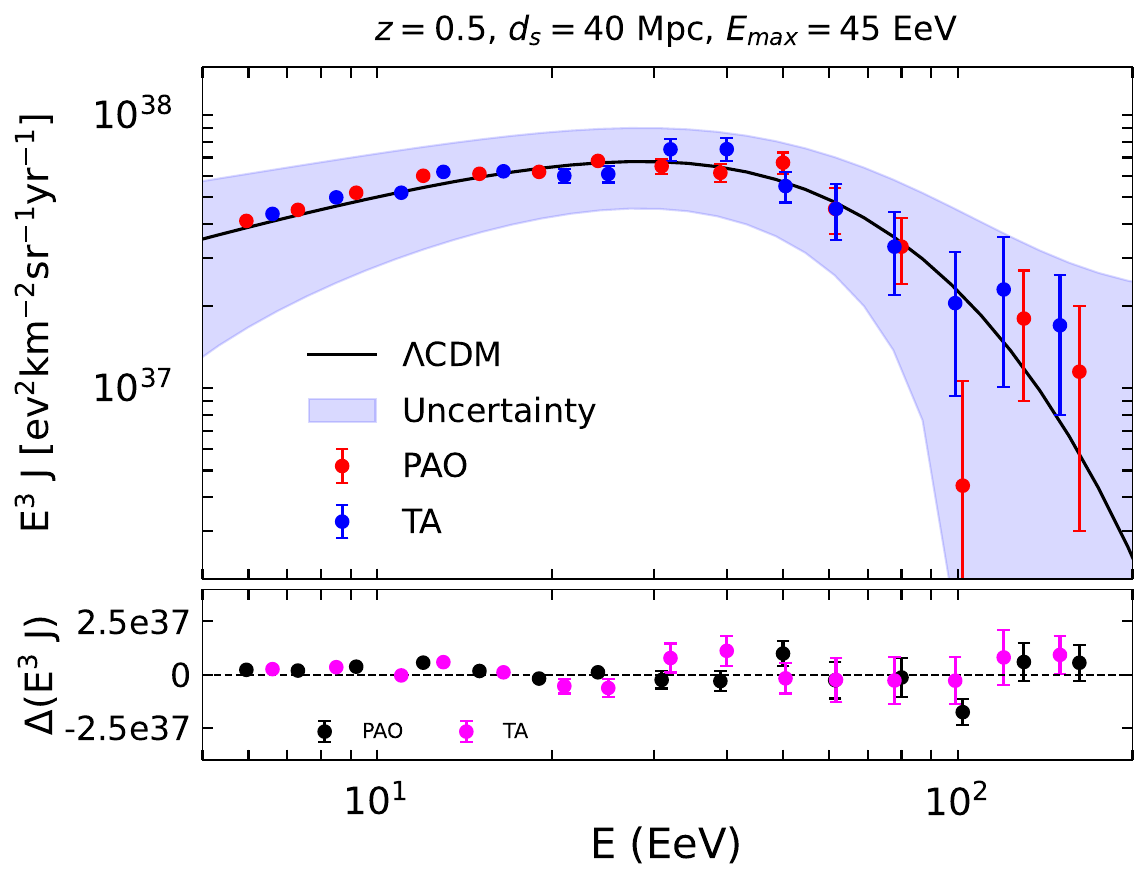}
\includegraphics[scale=0.28]{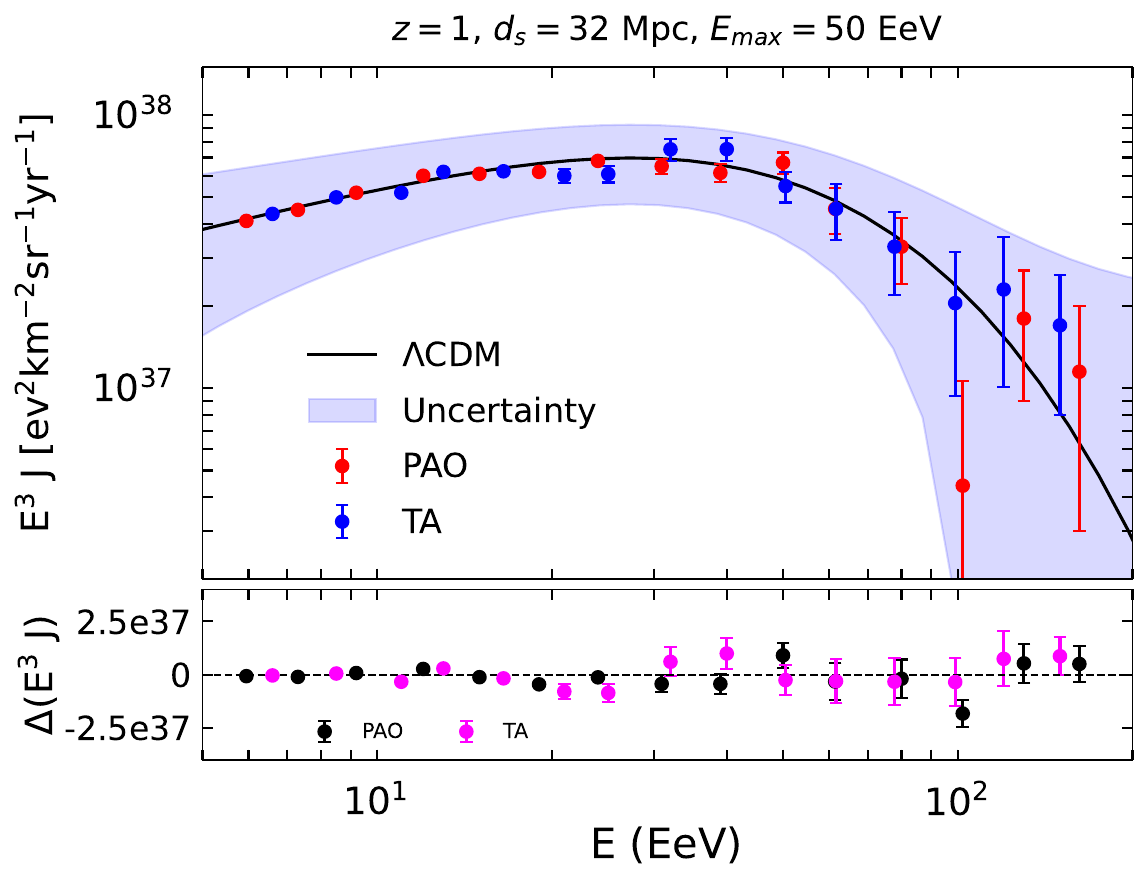}\hspace{5pt}
\includegraphics[scale=0.28]{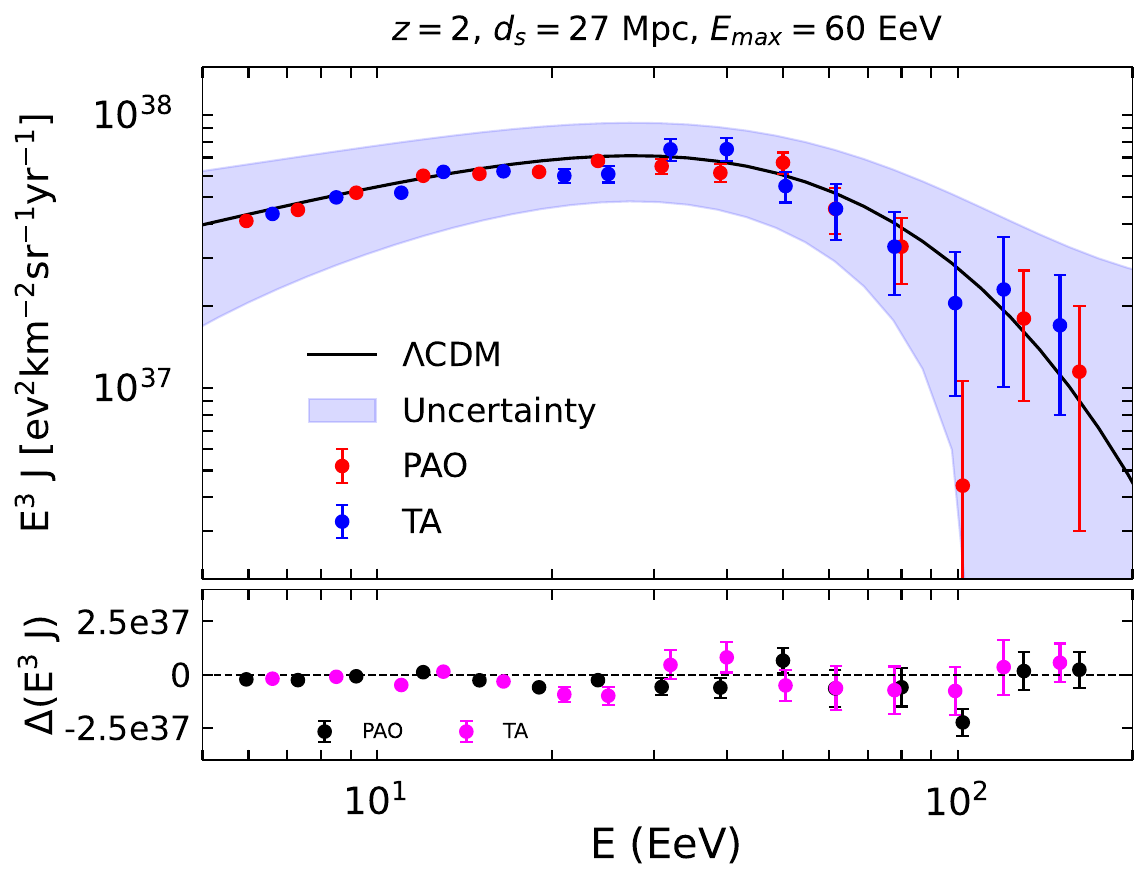}}
\vspace{-0.2cm}
\caption{UHECRs fluxes for different redshift $z$, separation distance of 
sources $d_\text{s}$, and $E_\text{max}$ for $f(R)$, $f(Q)$, and $\Lambda$CDM 
models. The observational data are from PAO \cite{augerprd2020} and 
TA \cite{ta2019}. The shaded regions depict the uncertainty in the flux 
predicted by the different cosmological models. Also, residue plots are shown 
highlighting the goodness of fitting.}
\label{flux}
\end{figure}

We calculate the UHECRs flux for the considered cosmological models through 
the magnetic field strength of $10$ nG. For this purpose, we take $150$ 
sources separated by a distance $d_\text{s}$. We perform a comparative 
analysis for the UHECR flux for the $f(R)$ gravity model, $f(Q)$ gravity 
model, and $\Lambda$CDM model in Fig.~\ref{flux_comp}. In the top panel of 
Fig.~\ref{flux_comp} we consider three different redshifts $z=0.5$, $1$, and 
$2$ and $d_\text{s}=30$ Mpc. Here, we also see that the model effect is 
significantly low in the lower redshift value. However, it is more pronounced 
in the higher redshift range. The redshift also shifts the spectrum up or down 
depending on its value. In the bottom panels of Fig.~\ref{flux_comp}, we fixed 
the redshift value and $d_\text{s}$, and we varied the $E_\text{max}$. We can 
see that the $E_\text{max}$ just changes the shape of the spectrum. It has no 
significant effect on the cosmological model, i.e.~$E_\text{max}$ effect is 
model independent. The shaded regions depict the uncertainty in flux from 
different cosmological models. Except the very first panel, the uncertainty predictions in Fig. \ref{flux_comp} are confined in the observational data range.
However, not all the cosmological models are fitted well with 
the PAO and TA data, we parametrise the $d_\text{s}$ with $z$ and 
$E_\text{max}$ in Fig.~\ref{flux}. Moreover, since the PAO and TA have 
different energy spectra, thus we have rescaled these data points as given 
in Ref.~\cite{bergmanepj} which is ${\Delta E}/{E} = \left[\pm 4.5 \pm 10 \log_{10} \left({E}/{10^{19} \text{ eV}} \right) \right]\%$.

\begin{table}[!h]
\caption{Parametrization of different parameters along with the $\chi^2$ 
values for $f(R)$, $f(Q)$, and the $\Lambda$CDM models.}
\vspace{5pt}
\centering
\begin{tabular}{c | c  | @{\hspace{7pt}}c@{\hspace{7pt}} | @{\hspace{7pt}}c@{\hspace{7pt}}| @{\hspace{7pt}}c@{\hspace{7pt}} | @{\hspace{7pt}}c@{\hspace{7pt}} | @{\hspace{7pt}}c | @{\hspace{7pt}}c | @{\hspace{7pt}}c | @{\hspace{7pt}}c }
Model & \hspace{5pt} z \hspace{5pt} & $d_\text{s}$ (Mpc) & $E_\text{max}$ (EeV) &  $\chi^2$ (PAO) & $\chi^2$ (TA) & $\chi^2$ (PAO + TA) & $\chi^2_{Red}$ (PAO) & $\chi^2_{Red}$ (TA) & $\chi^2_{Red}$ (PAO + TA)\\
\hline
 & 0.5 & 45 & 45 & 15.62 & 17.58 & 33.20 & 3.12 & 2.20 & 2.07\\

$f(R)$& 1 & 37 & 50 & 15.67 & 16.95 & 32.62 & 3.13 & 2.11 & 2.04  \\

 
 & 2 & 32 & 60 & 21.44 & 13.82 & 35.26 & 4.29 & 1.73 & 2.20  \\

\hline


& 0.5 & 40 & 45 & 14.96 & 20.17 & 35.13 & 2.99 & 2.52 & 2.19 \\

$f(Q)$& 1 & 30 & 50 & 12.34 & 11.15 & 23.49 & 2.47 & 1.39 & 1.47  \\

 
 & 2 & 24 & 60 & 13.81 & 10.82 & 24.63 & 2.76 & 1.35 & 1.54  \\
 \hline


& 0.5 & 40 & 45 & 12.52 & 10.83 & 23.35 & 2.50 & 1.35 & 1.46 \\

$\Lambda$CDM& 1 & 32 & 50 & 13.78 & 18.45 & 32.23 & 2.76 & 2.31 & 2.01 \\

 
 & 2 & 27 & 60 & 18.95 & 29.05 & 48.00  & 3.79 & 3.63 & 3.00\\
\hline
\end{tabular}
\label{tab1}
\end{table}


\begin{figure}[htb]
\centerline{
\includegraphics[scale=0.36]{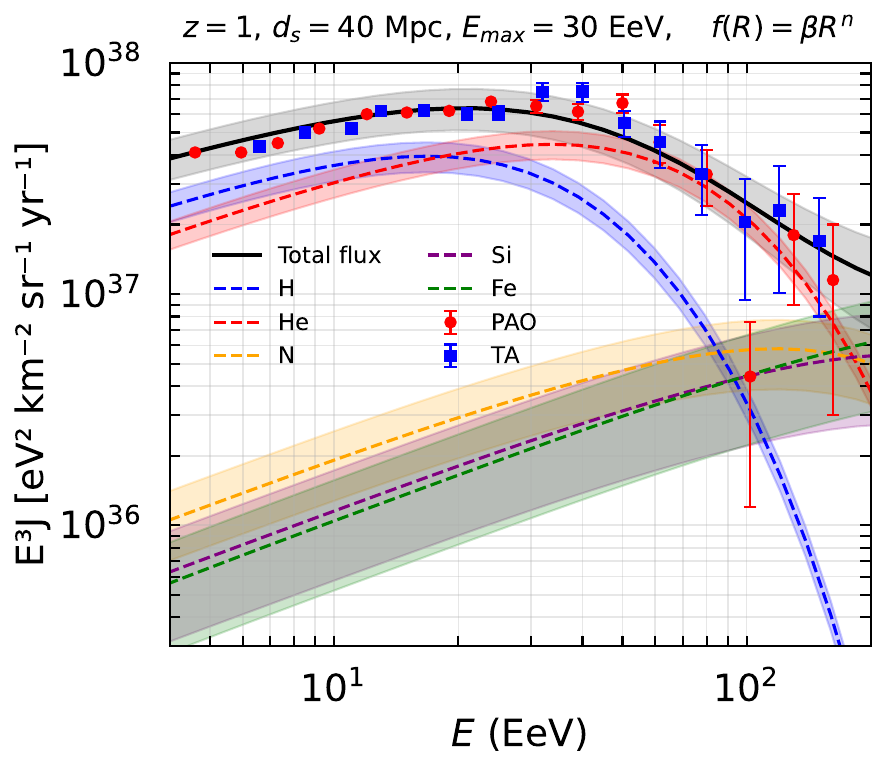}\hspace{15pt}
\includegraphics[scale=0.36]{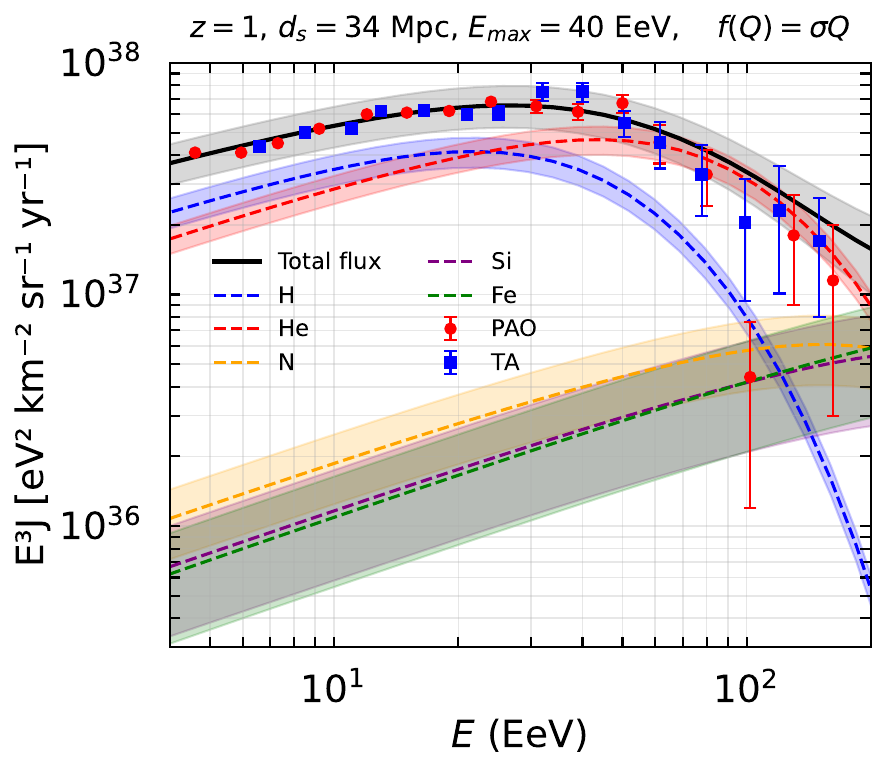}\hspace{15pt}
\includegraphics[scale=0.36]{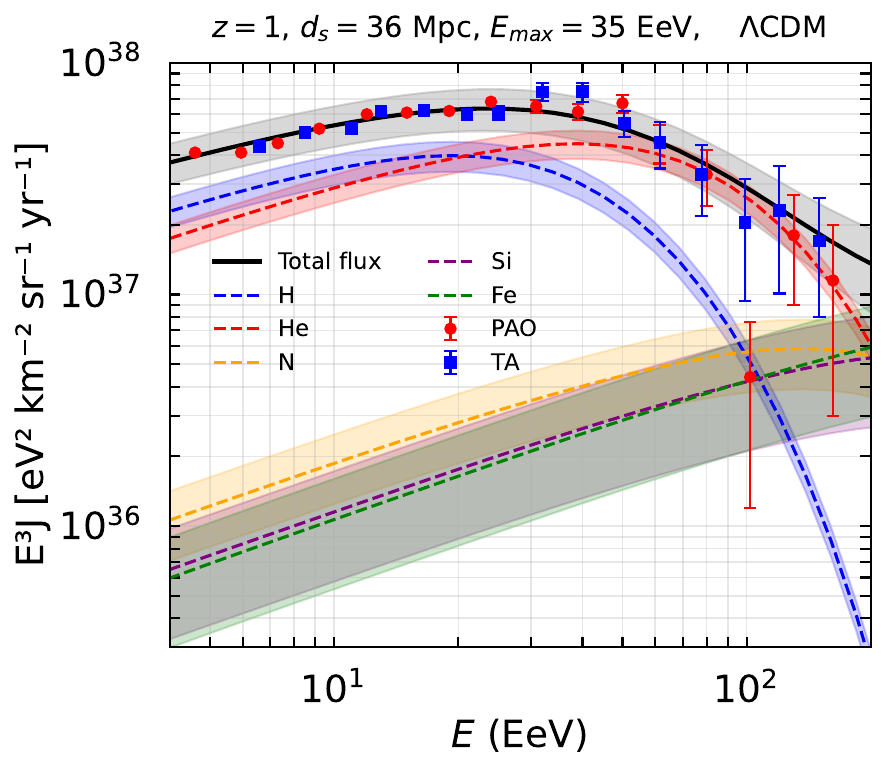}\hspace{15pt}}
\centerline{
\includegraphics[width=0.30\textwidth,height=0.20\textwidth]{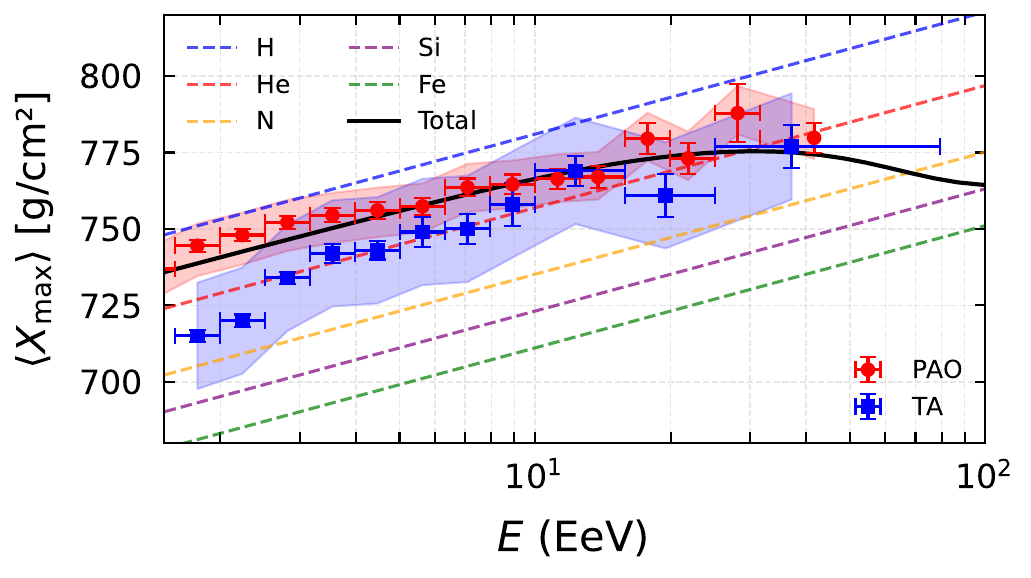}\hspace{17pt}
\includegraphics[width=0.30\textwidth,height=0.20\textwidth]{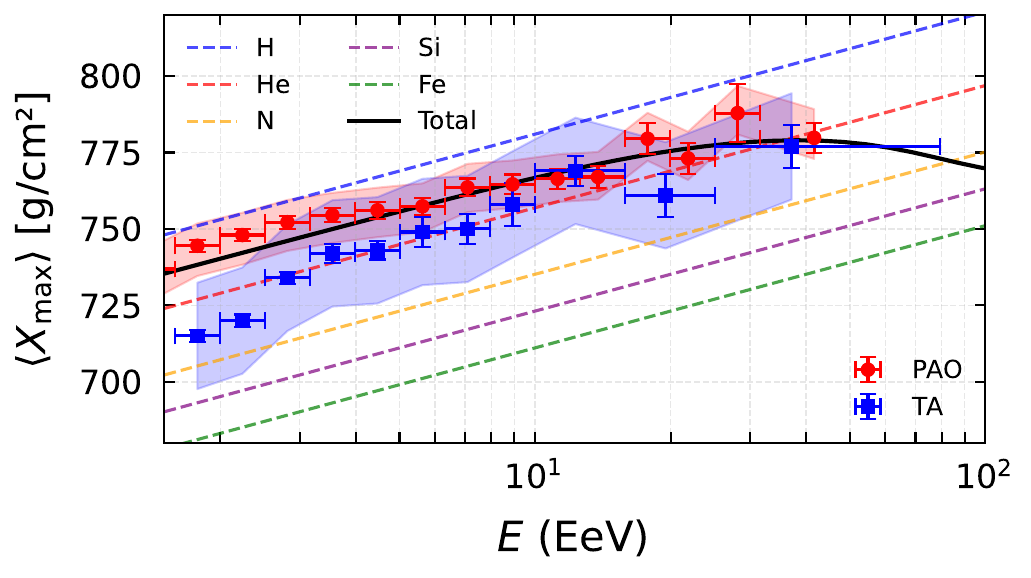}\hspace{15pt}
\includegraphics[width=0.30\textwidth,height=0.20\textwidth]{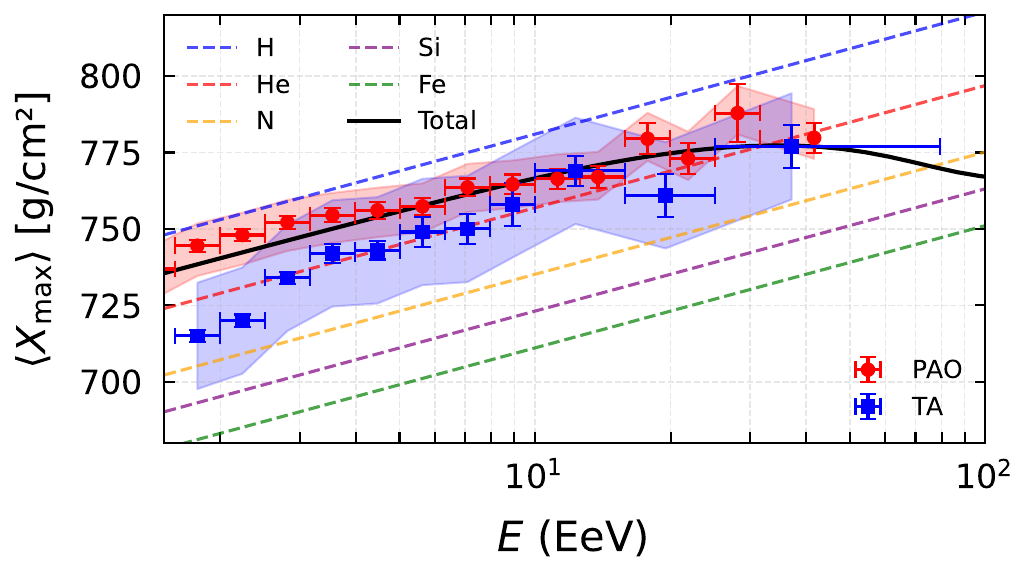}}\hspace{15pt}
\vspace{-0.4cm}
\caption{Upper: Mixed compositions of UHECR flux for $f(R)$, $f(Q)$, and $\Lambda$CDM model with the PAO and TA data. Lower: Corresponding $\langle X_{\text{max}} \rangle$ and compare with PAO and TA $\langle X_{\text{max}}\rangle $ data. }
\label{mixed}
\end{figure}

For this analysis, we take $E_\text{max}$ in such a way that
 $E< Z E_{max}/2$ \cite{moleachjcap} so that it gives a smooth  spectral 
shape (see Fig.~\ref{flux}). Three different values of $E_\text{max}$ have 
been taken into account for three different redshift values. We use the 
redshift values up to $2$ only, since the observational Hubble data has values 
for a maximum of $z=2.36$ till now. We consider the ranges of 
$z=0.5-2$, $d_\text{s}=24-45$ Mpc, and $E_\text{max}=45-60$ EeV. In this 
analysis, we set the same values of $E_\text{max}$ for each of the models, so 
that the fitting of the cosmological model will completely depend on the 
distance between the sources, the $d_\text{s}$ parameter only. For the 
goodness of fitting, we implement a residue plot for each of the cases 
considered, and one can see that our results have fitted very well with the 
data from PAO and TA. A list of parameterisations of the above parameters is 
given in Table \ref{tab1}. Also, we implement the $\chi^2$ test defined as
\begin{equation}
    \chi^2 = \sum_{i} \frac{(\text{J}^i_\text{th}-\text{J}^i_\text{obs})^2}{\sigma_i^2},
\end{equation}
where $\text{J}^i_\text{th}$ and $\text{J}^i_\text{obs}$ are the $i$th 
the theoretical value of flux obtained from our numerical calculations and 
experimental value of flux obtained from PAO and TA data, respectively. 
$\sigma_i$ is the error of the corresponding observed data. In Table 
\ref{tab1}, we provide the reduced $\chi^2$ value along with the $\chi^2$ 
value, which is obtained by dividing the $\chi^2$ value by its degree of 
freedom. In all cases, we provide the $\chi^2$ value for the individual 
PAO and TA data and also for their combination (PAO + TA). 
The shaded regions in Fig. \ref{flux} represent the uncertainties in predicting fluxes based on cosmological models considered here. These uncertainty bands are confined within the error bars of the observational data, indicating that the predicted flux ranges are consistent with the observed data.

 The fluxes discussed above are based on pure proton composition. However, 
any inference of physical parameters has to be made by using simultaneously 
both energy spectrum and composition data for better confidence on the basis 
of realistic situations. Thus, it is necessary to consider a mixed composition 
that reproduces the depth of shower maximum $X_\text{max}$. The 
$X_{\text{max}}$ is calculated using a parametrization derived from the air 
shower physics \cite{Auger2010, gaisser1990} for a mass number $A$ at energy 
$E$, and is given by \cite{Auger2010, gaisser1990}
\begin{equation}
X_{\text{max}}(E,A) = X_0 + \nu \ln\left(\frac{E}{A}\right),
\end{equation}
where $X_0$ and $\nu$ are coefficients that depend on the hadronic 
interaction process \cite{Auger2010}. In our calculations, we have taken 
these coefficients' values respectively as $700$ gm $\text{cm}^-2$ and 
$50$ gm $\text{cm}^-2$ \cite{gaisser1990, gaisser2016}. For the mixed 
composition scenarios, we computed a flux-weighted average 
$X_\text{max}$ as
\begin{equation}
\langle X_{\text{max}} \rangle = \frac{\sum\limits_i J_i(E) \cdot X_{\text{max},i}(E,A_i)}{\sum\limits_i J_i(E)},
\end{equation}
where $J_i(E)$ are the individual nuclear fluxes. In Fig.~\ref{mixed}, we 
plot the fluxes for a mixed composition of nuclei along with the 
corresponding $\langle X_{\text{max}} \rangle$ for $f(r)$, $f(Q)$, and 
$\Lambda$CDM model at $z=1$. It is seen that the parameterisations are 
different from the pure proton composition. The effect of the cosmological 
models is also visible here. It will be more pronounced if we plot for 
$z=2$ or higher. The observational $\langle X_{\text{max}} \rangle$ data 
(PAO and TA) are taken from Ref.~\cite{xmax_pao, xmax_ta}. The data contains 
$\langle X_{\text{max}} \rangle$ values, statistical errors, and statistical 
uncertainties. In the $\langle X_{\text{max}} \rangle$ plots, the statistical 
uncertainties are shown as bands. It can be seen that the obtained results are 
aligned with the observation, which depicts the validation of MTG and ATG on 
CRs spectra.

\section{Conclusion}\label{secVI}
The understanding of CRs propagation through galactic and extragalactic space 
has been a key research focus for several decades. It is suggested that the 
propagation of CRs through space may be significantly influenced by the 
existence of TMFs and the ongoing accelerated expansion of the Universe. 
This consideration motivates us for an exploration of CR propagation in TMFs 
within galactic and extragalactic space, framed within the context of $f(R)$  
and $f(Q)$ gravity theories, with a comparison to experimental data from two 
giant CR experiments: PAO and TA. We take into consideration two viable models: 
the $f(R)$ power-law model and a $f(Q)$ gravity model. Initially, the models' 
independent parameters are constrained using recent observational Hubble data. 
The relationship between redshift $z$ and evolution time $t$ is then computed 
for both models. We calculate the enhancement factor for galactic and 
extragalactic CRs within these two models' frameworks, along with the standard 
$\Lambda$CDM model. In Fig.~\ref{enhancement_comp1}, we perform a comparative 
analysis of the cosmological models in terms of CRs density enhancement factor, 
from where we got the effectiveness of the cosmological models in the higher 
redshift values. Fig.~\ref{fig_enhancement_z} illustrates the CRs density 
enhancement factor over an energy range of $0.01$ EeV to $100$ EeV, considering
cosmological redshifts from $0$ to $2.5$, using 100 bins in energy and 
redshift. We examine these scenarios for the source distances ($ d_\text{s} $) 
of $10$ Mpc, $20$ Mpc, $30$ Mpc and $40$ Mpc. The results indicate that at 
lower redshifts ($<1$), the density enhancement increases with energy up to 
approximately $5$ EeV before decreasing. At higher redshifts ($>1$), the 
density enhancement decreases with increasing energy up to $0.1$ EeV. Within 
the energy range of $0.1$ EeV to $5$ EeV, the enhancement factor decreases 
slowly with energy for $d_\text{s} = 10 $ Mpc. For $d_\text{s} = 20$ Mpc, the 
enhancement remains relatively flat across this energy range, while for 
$ d_\text{s} = 30 $ Mpc, a slow increase in enhancement is observed. At 
$ d_\text{s} = 40 $ Mpc, a clear increasing pattern is evident. This pattern 
suggests that as the distance between sources increases, the enhancement 
factor also rises. The enhancement factor remains significant up to 
approximately $70$ EeV, beyond which no further enhancement is observed. 
Comparing these results across three cosmological models, it is found that the 
$ f(R) $ model predicts the lowest enhancement, the $ f(Q) $ model predicts 
the highest, and the standard $ \Lambda \text{CDM} $ model shows moderate 
enhancement throughout the energy range. The effects of the cosmological 
models are more pronounced at higher source distances and redshifts, although 
no significant effects are observed for redshifts beyond $2.5$. In 
Fig.~\ref{enhancement_comp2}, a comparative analysis has also been performed 
within the cosmological framework by varying the magnetic field strength and 
we get that the higher magnetic field has no significant effect. 
Fig.~\ref{fig_enhancement_b} shows the enhancement factor for  CRs across 
different magnetic field amplitudes, with the redshift set to $ z=1 $, the 
source distance $ d_\text{s}  = 10$ Mpc, and the coherence length 
$ l_\text{c} =1$ Mpc. In both high and low energy regions, the $ f(Q) $ model 
predicts the highest enhancement, followed by the $ \Lambda \text{CDM} $ and 
$ f(R) $ models. At lower energy ranges, the effects of the cosmological model 
are noticeable across all magnetic field strengths, while at higher energies, 
the magnetic field's contribution to CRs enhancement diminishes. The bottom 
panels of Fig.~\ref{fig_enhancement_b} depict the same analysis with a source 
distance of $ d_\text{s} = 40$ Mpc. Here, the magnetic field significantly 
affects the enhancement factor up to $10$ EeV, after which the contributions 
from the magnetic field become uniform. The $f(R)$ model enhances the 
amplitude from about $10$ to $9000$, whereas the $ f(Q) $ model provides the 
highest overall enhancement. This demonstrates that cosmological models 
significantly influence CR propagation depending on redshift, magnetic field 
strength, source distance, and energy.

A simple comparative analysis of cosmological models for the CRs enhancement 
factor in the case of nuclei has been done in Fig.~\ref{enhancement_comp3}, and 
we find that in both low and high energy ranges, the heavy nuclei give a 
better cosmological effect as compared to the lighter ones. In 
Fig.~\ref{fig_enhancement_nuclei}, we analyse the enhancement factor for 
different nuclei (helium, nitrogen, and iron) with a magnetic field strength 
of $10$ nG. For helium nuclei, the $ f(Q) $ model predicts the highest 
enhancement in the energy range of $0.01$ to $10$ EeV. For energies above 
$10$ EeV, the predictions across all cosmological models converge. Similar 
results are observed for nitrogen and iron nuclei, although the density 
enhancement for iron nuclei increases gradually after reaching specific energy 
thresholds for each model.

We also compute and compare the UHECRs fluxes for these cosmological models 
with a magnetic field strength of $10$ nG, using $150$ sources separated by 
$ d_\text{s} $ in Fig.~\ref{flux_comp}. We see that the cosmological model 
effect occurs by $z$ only, while $E_\text{max}$ changes the shape of the 
spectrum depending on its value. We fit the models' predictions to 
observational data from the PAO and TA by adjusting parameters $z$, 
$ d_\text{s} $, and $ E_\text{max} $. Redshift values are limited to $2$, 
given the current observational constraints. We test ranges of $ z = 0.5 - 2 $,
 $ d_s = 24 - 45 $ Mpc, and $ E_\text{max} = 45 - 60 $ EeV, keeping 
$ E_\text{max} $ constant across models to isolate the impact of 
$ d_\text{s} $. We found that at $z=0.5$, the best fitting value for 
$d_\text{s}$ are $45$ Mpc, $40$ Mpc, and $40$ Mpc for the $f(R)$, $f(Q)$, and 
the $\Lambda$CDM model respectively. Similarly, for $z=1$, these values of 
$d_\text{s}$ are $37$ Mpc, $30$ Mpc and $32$ Mpc, and for $z=2$, these are 
$32$ Mpc, $24$ Mpc, and $27$ 
Mpc respectively. Residual plots show good agreement with PAO and TA data. 
Table~\ref{tab1} also presents the $ \chi^2 $ values as well as the reduced 
$\chi^2$ for individual PAO and TA data, as well as their combined data 
(PAO + TA). So from Fig.~\ref{flux_comp} and \ref{flux}, we can conclude 
that $d_\text{s}$ shifts the whole spectrum up and down depending on the 
amplitude. The values of redshift $z$ show the effectiveness of cosmological 
models in the energy spectrum and shift the spectrum, while $E_\text{max}$ 
plays the role in the changing of spectrum shape at $E<Z E_\text{max}/2$. 
Furthermore, in Fig.~\ref{flux}, the uncertainty plot with PAO and TA data 
supports the validation of the considered MTG and ATG models in UHECRs domain.
The UHECR flux for the mixed composition and respective 
$\langle X_{\text{max}} \rangle$ plot are shown in Fig.~\ref{mixed}. In this
case also the obtained results are also aligned with the 
$\langle X_{\text{max}} \rangle$ data of PAO and TA.

The analysis highlights the significant role that cosmological models and 
magnetic fields play in the propagation and enhancement of CRs. The $f(Q)$ 
model consistently predicts the highest enhancement, suggesting that ATGs may 
offer a framework for understanding CR behaviour in the Universe. Our obtained 
results of UHECRs flux are consistent with observations from the PAO and 
TA, supporting the validity of these models in explaining CRs behaviour.
It is to be noted that in some cases, the cosmological effect is less 
pronounced.
This is because we use the redshift $z$ up to $2.5$, 
based on the availability of observational data. The cosmological effect 
would become more significant if higher redshift values were considered.
Moreover, we use the UHE range in our entire work, but extending 
to lower energies could better reveal the differences in adiabatic losses and 
thus may be larger cosmological differences too,ß due to different gravity 
models.

\section*{Acknowledgements} 
UDG is thankful to the Inter-University Centre for Astronomy and Astrophysics 
(IUCAA), Pune, India, for the Visiting Associateship of the institute.
The authors gratefully acknowledge the anonymous reviewer for his/her valuable comments and suggestions.

\end{document}